\documentclass[pra,aps, twocolumn,groupedaddress,showpacs]{revtex4-1}
\usepackage{amssymb,amsmath,amsfonts,epsfig,graphicx,latexsym,bm,color,braket}

\begin{document}

\title{Bosons condensed in two modes with flavour-changing interaction}

\author{Andreas Hemmerich\footnote{e-mail: hemmerich@physnet.uni-hamburg.de}}
\affiliation{Institut f\"ur Laser-Physik and Hamburg Center of Ultrafast Imaging, Universit\"at Hamburg, D-22761 Hamburg, Germany}
\date{\today}

\begin{abstract}
A quantum model is considered for $N$ bosons populating two orthogonal single-particle modes with tunable energy separation in the presence of flavour-changing contact interaction. The quantum ground state is well approximated as a coherent superposition (for zero temperature) or a mixture (at low temperature) of two quasi-classical states. In a mean field description, the systems realizes one of these states via spontaneous symmetry breaking. Both mean field states, in a certain parameter range, possess finite angular momentum and exhibit broken time-reversal symmetry in contrast to the quantum ground state. The phase diagram is explored at the mean-field level and by direct diagonalisation. The nature of the quantum ground state at zero and finite temperature is analyzed by means of the Penrose Onsager criterion. One of three possible phases shows fragmentation on the single-particle level together with a finite pair order parameter. Thermal and quantum fluctuations are characterized with respect to regions of universal scaling behavior. The non-equilibrium dynamics shows a sharp transition between a self-trapping and a pair-tunneling regime. A recently realized experimental implementation is discussed with bosonic atoms condensed in the two inequivalent energy minima $X_{\pm}$ of the second band of a bipartite two-dimensional optical lattice. 
\end{abstract}


\maketitle

\section{Introduction}
The vast complexity of natural many-body systems often leads us to seek ab initio tractable minimal quantum-mechanical models, which can capture a few isolated phenomena of interest while excluding the superimposed jungle of secondary structure that would impede a clear understanding. Such model systems often find their experimental counterpart in ultracold quantum gases and more specifically atomic Bose-Einstein condensates (BECs) \cite{Leg:01, Pet:02, Pit:03} if bosonic particles are of interest, as in these notes. The restriction to only a few or even to only two single-particle modes significantly simplifies things, but still leaves room to capture relevant physics such as that of Josephsen dynamics \cite{Sme:97, Mil:97, Gri:98, Leg:01, Dal:12}. Well known simple interacting many-body Hamiltonians with only two single-particle modes are the Nozi$\grave{\textrm{e}}$res model \cite{Noz:95} $H^{(1)} = \frac{g}{2} n_{1} n_{2}$, with particle numbers $n_{i} = a_{i}^{\dagger} a_{i}$, annihilation operators $a_{i}, i \in \{1,2\}$ for states $\ket{1}, \ket{2}$ and an either repulsive ($g>0$) or attractive ($g<0$) interaction, or the two-site bosonic Hubbard model of $N$ bosons tunneling in a double well given by $H^{(2)} = -t (a_{1}^{\dagger} a_{2}+a_{2}^{\dagger} a_{1})  +  \frac{U}{2} (n_{1} (n_{1}-1)+n_{2} (n_{2} - 1))$, comprising a tunneling term with typically positive tunneling strength $t$ and an on-site collision term with collision energy per particle $U$ \cite{Ruo:98, Kal:01, Mue:06, Sal:07}. Despite their simplicity, these models, depending on the external parameters, show different fragmented ground states \cite{Mue:06} and non-linear dynamics that can give rise to a suppression of tunneling referred to as self-trapping \cite{Sme:97, Mil:97, Rap:12}. Experimental implementations of non-linear two-mode dynamics including self-trapping have been reported with superfluid helium \cite{Bac:98}, atomic BECs \cite{Alb:05, Lev:07, Foe:07}, or condensates of exciton-polaritons \cite{Abb:13}.

In this article a bosonic two-mode model is considered with a more general interaction, consisting of the following three parts: a term proportional to $n_{1} n_{2}$ as in the Nozi$\grave{\textrm{e}}$res model, an on-site term $n_{1} (n_{1}-1)+n_{2} (n_{2} - 1)$ as in the two-site bosonic Hubbard model and, most importantly, a flavour-changing interaction $a_{1}^{\dagger} a_{1}^{\dagger} a_{2} a_{2} + a_{2}^{\dagger} a_{2}^{\dagger} a_{1} a_{1}$, which describes two atoms colliding in one of the modes with the result that both particles are transferred to the other mode. Such flavour changing interactions typically arise in scenarios where orbital degrees of freedom provide degeneracies. For example, the two modes could be $p_x$- and $p_y$-orbitals in the first excited state of a two-dimensional (2D) harmonic oscillator. The flavour-changing character of the interaction mimics a pair tunneling term between the two single-particle modes \cite{Lia:09, Bad:09, Cao:12, Jas:12, Fis:13, Zhu:15, Rub:17, Agb:18} and as such it should act to induce coherence between these modes, in contrast to flavour conserving interactions, which tend to inhibit coherence \cite{Mue:06}. It is found that the quantum ground state is well approximated as a coherent (for zero temperature) or incoherent (at low temperature) superposition of two quasi-classical phase states, each of which can be realized in a mean-field description via spontaneous symmetry breaking. The two phase states are well approximated by a macroscopically populated superposition of the two single-particle modes with a phase difference of either $\pi/2$ or $-\pi/2$ and hence possess finite angular momentum and exhibit broken time-reversal symmetry. This contrasts with the quantum ground state of the system, for which the expectation value of the angular momentum is zero, however with large fluctuations on the order of $\hbar$ per particle. The phase diagram is explored at the mean-field level and by direct diagonalisation. The nature of the quantum ground state at zero and finite temperature is analyzed by means of the Penrose Onsager criterion \cite{Pen:56}. When both single-particle modes are populated, the ground state shows fragmentation on the single-particle level together with a finite pair order parameter. Thermal and quantum fluctuations are characterized with respect to regions of universal scaling behavior. Note that some of the findings for zero temperature were previously discussed in Refs.~\cite{Bad:09, Fis:13, Zhu:15}. The non-equilibrium dynamics shows a sharp transition between a self-trapping regime, where the atoms initially prepared in a single mode remain in that mode, and a pair-tunneling regime, in which the atoms perform Josephson oscillations. A recently realized experimental implementation is discussed with bosonic atoms condensed in the two inequivalent energy minima $X_{\pm}$ of the second band of a bipartite 2D optical lattice. 

\section{Model}
The general Hamiltonian for bosons subject to binary contact interaction $H = \int d^3r ( \psi^{\dagger} H_0 \psi + \frac{g}{2} \psi^{\dagger} \psi^{\dagger} \psi \psi )$ is considered, with the single-particle Hamiltonian $H_0$ and the collision parameter $g>0$. The bosonic field operator $\psi(r)$ is decomposed with respect to a basis set of single-particle modes $\ket{\alpha_{n}}$ according to $\psi(r) = \sum_n \,\alpha_n(r)\, a_{n}$, where $\alpha_{n}(r) \equiv \bra{r} \alpha_{n}\rangle$, $\ket{r}$ denotes the position basis, the normalization $\int d^3r \,\alpha_n \alpha_m^{\ast} = \delta_{nm}$ holds, and $a_{n}$ denote bosonic anihilation operators satisfying the commutation relations $[a_{n},a_{m}^{\dagger}] = \delta_{nm}$. It is easily verified that $[\psi(r),\psi^{\dagger}(r')]$ = $\sum_{n} \alpha_{n}(r) \alpha_{n}^{\ast}(r')$ = $\bra{r}r'\rangle$ = $\delta(r-r')$. Now, assume that two of the single-particle modes, namely $\alpha_1(r)$ and $\alpha_2(r)$, are exclusively populated and not coupled to any others and that these two modes are eigenmodes of the single-particle Hamiltonian $H_0$, thus fulfilling the relations $H_0 \alpha_i = \varepsilon_i \alpha_i, i \in \{1,2\}$ with energy eigenvalues $\varepsilon_i = (-1)^{i} \frac{1}{2}\varepsilon $. This leads to the  Hamiltonian
\begin{eqnarray}
\label{eq:Hamiltonian}
H &=& \frac{\varepsilon}{2} (n_{2}-n_{1}) + \frac{g}{2} \left[\rho_{0,1}\, n_{1} (n_{1}-1) + \rho_{0,2}\, n_{2} (n_{2} - 1)\right] \nonumber \\
&+&  2 g \,\rho_{1} n_{1} n_{2}  +  \frac{g}{2} \, (\rho_{2}^{\ast} \, a_{1}^{\dagger} a_{1}^{\dagger} a_{2} a_{2} + \rho_{2} \, a_{2}^{\dagger} a_{2}^{\dagger} a_{1} a_{1})
\end{eqnarray}
with the collision integrals $\rho_{0,1} \equiv \int d^3r |\alpha_1| ^4$, $\rho_{0,2} \equiv \int d^3r |\alpha_2|^4$,  $\rho_{1} \equiv \int d^3r |\alpha_1| ^2 |\alpha_2|^2$,  and $\rho_{2} \equiv \int d^3r \alpha_1^{\ast} \alpha_1^{\ast} \alpha_2 \alpha_2$. Note that because $\alpha_i$ are orthogonal one may approximate $\rho_{3} \equiv \int d^3r |\alpha_1| ^2 \alpha_1^{\ast} \alpha_2 \approx 0$ and $\rho_{4} \equiv \int d^3r |\alpha_2| ^2 \alpha_1^{\ast} \alpha_2 \approx 0$. Such processes would correspond to collisions of two particles in one of the modes leading to a transfer of only one of those particles to the other mode. Energy momentum conservation often entirely prevents such processes. In the concrete experimental implementation, discussed below, this approximation is very well fulfilled. In the following, it is assumed that both modes are associated with the same on-site collision energy per particle, i.e., $\rho_{0} \equiv \rho_{0,1} = \rho_{0,2}$. This implies $\rho_{1} \leq \rho_{0}$ as an immediate consequence of $(|\alpha_1|^2 - |\alpha_2|^2)^2 \geq 0$. The Hamiltonian $H$ obviously includes the Nozi$\grave{\textrm{e}}$res model $H^{(1)}$ but also the two-site Hubbard model $H^{(2)}$. The latter is seen by rewriting $H^{(2)}$ with respect to the eigenbasis of its tunneling term. The following discussion is further simplified by assuming time-reversal symmetry of $H_0$ and hence real mode functions $\alpha_i(r)$, such that $\rho_{1} = \rho_{2}$. Furthermore, the collision parameters $g_0 \equiv g \rho_0$ and $g_1 \equiv g \rho_1 = g \rho_2$ are employed. According to the constraints for $\rho_0$ and for $\rho_1$ found above, one has $g_1/g_0 \in [0,1]$. With these simplifications one may write the Hamiltonian in Eq.~\ref{eq:Hamiltonian} in terms of the operator $L \equiv \frac{1}{N}i (a_1 a_2^{\dagger}-a_1^{\dagger} a_2 )$, which may be interpreted as an orbital angular momentum per particle in units of $\hbar$ \cite{Isa:05, Liu:06}. This leads to 
\begin{eqnarray}
\label{eq:Hamiltonian2}
H &=& \frac{\varepsilon}{2}\, (n_{2} - n_{1}) + \frac{g_{0}}{2} \,  \left[n_{1} (n_{1}-1) + n_{2} (n_{2} - 1)\right]  \nonumber \\
&+& \, \, \, \, \, \frac{g_1}{2}\,  (N + 6 \,n_{1} n_{2})  -\frac{g_{1}}{2}\,  N^2 \,L^2 \,.
 \end{eqnarray}
Note that the angular momentum term is negative if $g_{1}>0$, such that the system should have a tendency to maximize $L^2$ in order to minimize its energy.
\begin{figure}
\includegraphics[scale=0.5, angle=0, origin=c]{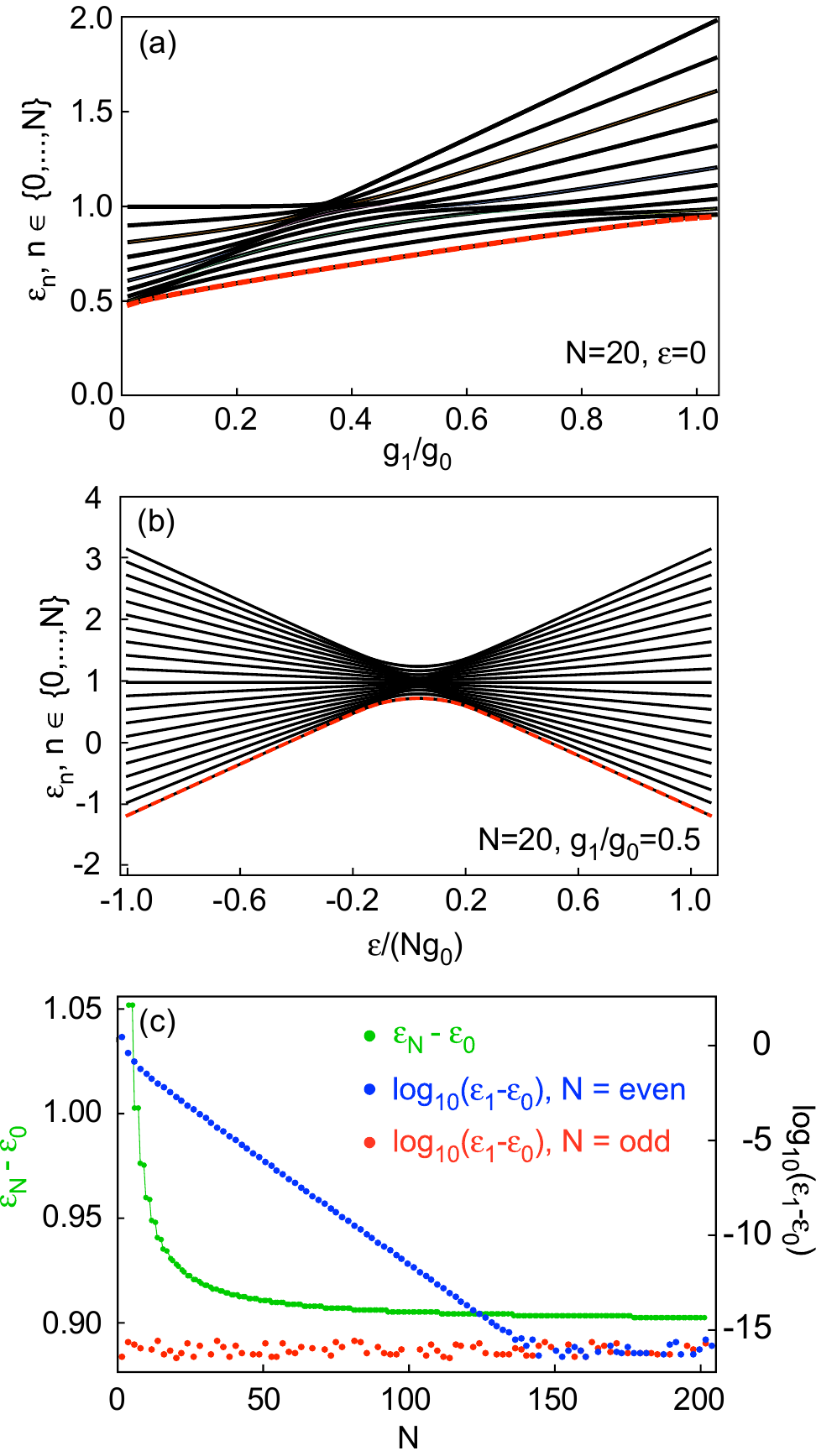}
\caption{The eigenenergies $\varepsilon_n, n \in \{0,...,N\}$ of $H$ for $N=20$ are plotted versus $g_1/g_{0}$ for $\varepsilon=0$ in (a) and versus $\varepsilon/(Ng_0)$ for $g_1/g_{0}=0.5$ in (b). The thick red dashed lines highlight the ground state energy $\varepsilon_0$. (c) Plots of $\log_{10}(\varepsilon_1-\varepsilon_0)$ (red disks for odd $N$, blue disks for even $N$) and $\varepsilon_N - \varepsilon_0$ (green trace) versus particle number for $\varepsilon=0$ and $g_1/g_{0} = 0.9$. All plots do not depend on the value of $g_{0}$.}
\label{fig:Fig.1}
\end{figure}

\section{Structure of energy eigenstates}
In order to study its eigenvalues $E_n$ and eigenstates $\ket{E_n}$, $H$ is straight forwardly diagonalised in the Fock-basis $\ket{\nu} \equiv \ket{N - \nu, \nu}$ with $\nu \in \{0, ..., N\}$, which consists of the states with exactly $N-\nu$ atoms in mode $\alpha_1$ and $\nu$ atoms in $\alpha_2$. Our numerical code lets us employ values of $N$ exceeding several times $10^{4}$, i.e., sufficiently large to match with experimental implementations, as discussed later. It is found that the eigenstates $\ket{E_n}$ are superpositions of Fock states $\ket{\nu}$ with either only even or only odd values of $\nu$, which may be interpreted as a direct consequence of the interaction-induced pairwise exchange of particles between the two single-particle modes. In Fig.~\ref{fig:Fig.1}(a) the eigenvalues $E_{n}, n \in \{0,N\}$ of $H$ (indexed in ascending order) are shown as $g_1/g_{0}$ is tuned across the interval $[0,1]$ for fixed $g_{0}$ and $\varepsilon=0$. The normalized energy $\varepsilon_n \equiv E_{n}/ W_0$ is plotted, accounting for the fact that the relevant energy span of the eigenenergies scales with the on-site collision energy associated with each mode $W_0 \equiv g_0 N(N-1)/2$. In (b) $g_1/g_{0} = 0.5$ is chosen while $\varepsilon$ is tuned across the interval $[-1,1] \times g_0 N$. For optimal visibility, a relatively small value $N=20$ is used. In (a) and (b), for fixed $g_1$ and $\varepsilon$, the values of $\varepsilon_n$ cover the same regions on the y-axes regardless of the choice of $g_{0}$ or $N$, although at an increased density of states if $N$ is increased. In (a) a pronounced resonance of the density of states becomes visible at an energy that linearly increases from 1/2 to 1 as $g_1/g_0$ is tuned from 0 to 1/3 and then remains at 1 for $1/3 < g_1/g_0 < 1$. This resonance plays an important role for the dynamical properties of the system as is discussed below. For $\varepsilon = 0$ and odd $N$ there are $(N+1)/2$ doubly degenerate eigenstates. Also for even $N$, the energies close to the upper and lower boundary of the energy spectrum tend to approximately arrange in nearly degenerate pairs with energy separations decreasing exponentially fast with $N$ (see also the discussion below Eq.(4) in Ref.~\cite{Bad:09}). This is seen in Fig.~\ref{fig:Fig.1}(c), where the logarithm of the energy difference between the two lowest eigenstates in units of $W_0$ is plotted versus $N$ (blue disks for even $N$, red disks for odd $N$). The noisy floor, found for all odd $N$ and those even $N$ exceeding about 140, represents the machine precision of the calculation. In the green trace, one also sees that the width of the eigenenergy spectrum divided by $W_0$ rapidly approaches a constant near unity for large $N$. For all traces of (c) $g_1/g_{0} = 0.9$ and $\varepsilon=0$. 

\section{Ground state at finite temperature}
A set of particularly useful states, playing a significant role in the discussion of the ground state of $H$, are the two-mode coherent states or - more briefly - phase states $\ket{\phi,\theta} \equiv \frac{1}{\sqrt{N}}(\cos(\theta) a_1^{\dagger}+ \sin(\theta) e^{i\phi} a_2^{\dagger})^N \ket{vac}$ with $\ket{vac}$ denoting the vacuum \cite{Mue:06}. Their projections onto the Fock basis read 
\begin{eqnarray}
\label{phasestate}
\langle \nu | \phi,\theta \rangle = \cos^{N-\nu}(\theta) \sin^{\nu}(\theta) \sqrt{{N \choose \nu}} \, e^{i\nu \phi} \, .
\end{eqnarray}
By virtue of their construction, these states possess a well defined relative phase $\phi$ between the two single-particle modes $\alpha_i, i = 1,2$, in the sense that the two sub-samples of particles belonging to $\alpha_i$ exhibit maximal mutual coherence. The angle $\theta$ determines the mean particle numbers in the two modes $\alpha_i$ as $\bra{\phi,\theta} n_1 \ket{\phi,\theta} = N \cos^2(\theta)$ and $\bra{\phi,\theta} n_2 \ket{\phi,\theta} = N \sin^2(\theta)$. For the case of equal mean particle numbers in both modes, i.e., $\theta = \pi/4$, the short notation $\ket{\phi} \equiv \ket{\phi,\pi/4}$ is used. In the following discussion, primarily the phase states $\ket{\pm \pi /2}$ and their superposition with arbitrary relative phase $e^{i2 \pi z}$, i.e., the \textit{cat} state $\ket{cat(z)} \equiv (\ket{\pi / 2} \, +  \, e^{i2 \pi z} \ket{-\pi / 2}) / \sqrt{2}$ are of interest. Their Fock basis coefficients are
\begin{eqnarray}
\label{eq:phaseandcat}
\langle \nu | \pm \pi /2 \rangle = \frac{(\pm i)^{\nu}}{\sqrt{2^N}} \sqrt{{N \choose \nu}} , \nonumber \\ \\ \nonumber
\langle \nu | cat(z) \rangle = \frac{ i^{\nu} (1+(-1)^{\nu} e^{i \pi z}) }{\sqrt{2^{N+1}}} \sqrt{{N \choose \nu}}.
\end{eqnarray}
\begin{figure}
\includegraphics[scale=0.2, angle=0, origin=c]{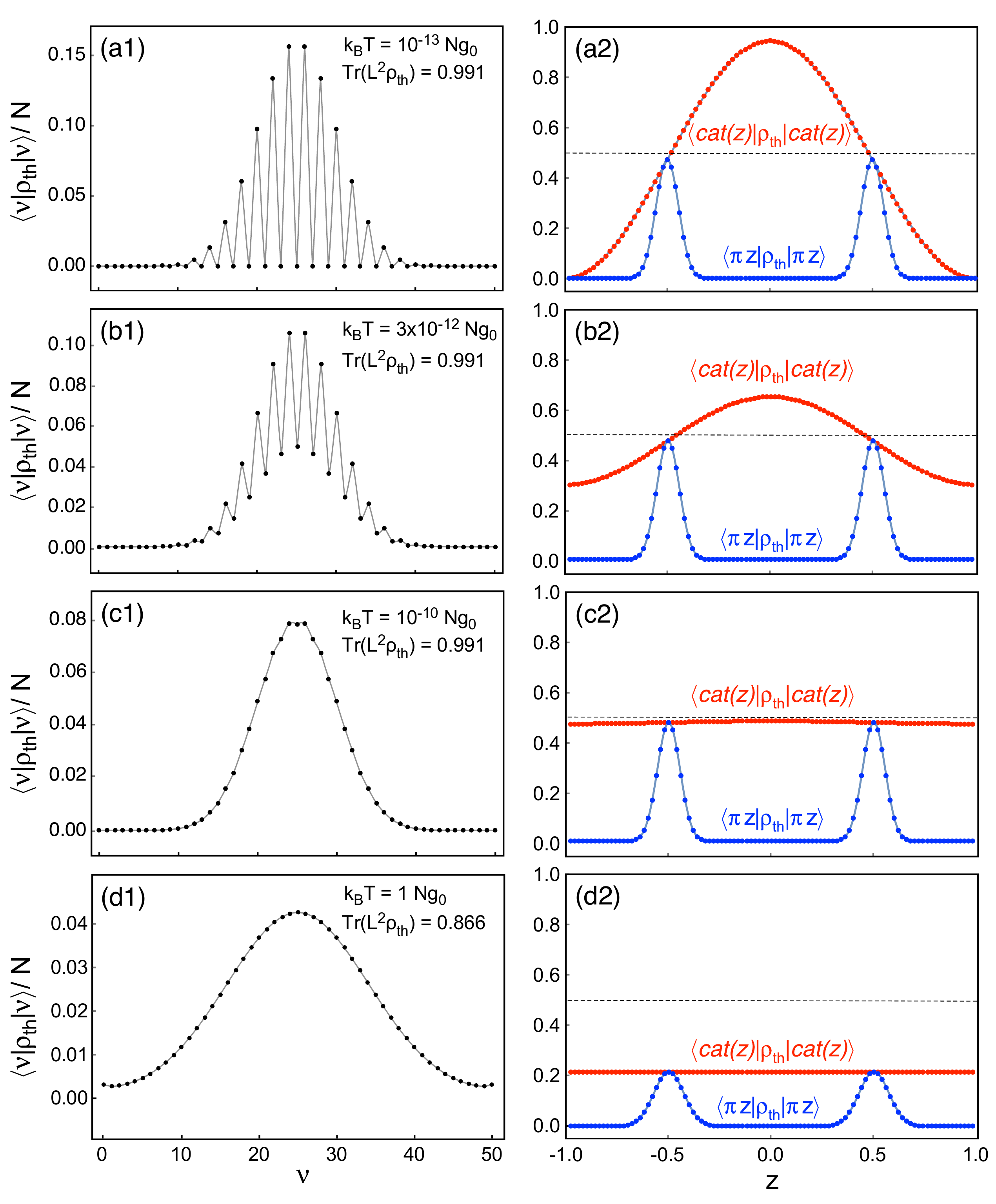}
\caption{For $\varepsilon = 0$ and $g_1 = \frac{2}{3} g_0$, (a1 - d1) show the populations $\bra{\nu}\rho_{th}\ket{\nu}$ for increasing temperature. The corresponding projections $\bra{cat(z)}\rho_{th}\ket{cat(z)}$ (red traces) and $\bra{\pi z}\rho_{th}\ket{\pi z}$ (blue traces) are plotted in (a2 - d2). In all plots $N=50$.}
\label{fig:Fig.2}
\end{figure}
The physical properties of these states can be characterized in terms of the angular momentum operator $L$. A straight forward calculation yields the expectation values $Tr(L\ket{\pm \pi /2} \bra{\pm\pi/2}) = \pm 1$, $Tr(L^2 \ket{\pm \pi /2} \bra{\pm\pi/2}) = 1$ and for the incoherent superposition of phase states $Tr(L \frac{1}{2}[\ket{\pi/2} \bra{\pi/2}+\ket{-\pi/2} \bra{-\pi/2}]) = 0$ and $Tr(L^2 \frac{1}{2}[\ket{\pi/2} \bra{\pi/2}+\ket{-\pi/2} \bra{-\pi/2}]) = 1$. Similarly, for the cat state $\ket{cat(z)}$ one gets $Tr(L \ket{cat(z)}\bra{cat(z)}) = 0$ and $Tr(L^2 \ket{cat(z)}\bra{cat(z)}) = 1$. Equipped with these remarks, one may explore the finite temperature ground state of $H$. To this end, the density operator 
\begin{eqnarray}
\label{eq:thermalstate}
\rho_{th} \equiv \frac{1}{Z} \sum_{n=0}^{N} e^{-\frac{E_{n}}{k_B T}}\ket{E_{n}}\bra{E_{n}}, Z \equiv \sum_{n=0}^{N} e^{-\frac{E_{n}}{k_B T}}
\end{eqnarray}
is used to calculate the populations in the Fock basis $\bra{\nu}\rho_{th}\ket{\nu}$, and the projections $Tr( \rho_{th} \ket{\pi z}\bra{\pi z})$ and $Tr( \rho_{th}\ket{cat(z)}\bra{cat(z)})$ with respect to the phase state $\ket{\phi}=\ket{\pi z}$ and the cat state $\ket{cat(z)}$ for $z$ tuned across the interval $[-1,1]$. Here, $\varepsilon = 0$ and $g_1 = \frac{2}{3} g_0$ is chosen, which falls within the range accessible in experiments, as discussed below. The results are shown in Fig.~\ref{fig:Fig.2} for four different temperatures. In the uppermost row the temperature is set to be practically zero ($k_B T = 10^{-13} Ng_0$). Since an even particle number $N=50$ is chosen, according to Fig.~\ref{fig:Fig.1}(c) the lowest energy eigenstate $\ket{E_{0}}$ is truly non-degenerate, separated from the first excited energy eigenstate by a tiny energy gap, which exponentially decreases with $N$ but exceeds $k_B T$. Hence, only $\ket{E_{0}}$ notably contributes to $\rho_{th}$. As seen in Fig.~\ref{fig:Fig.2}(a1), this leads to a characteristic form of the populations $\bra{\nu}\rho_{th}\ket{\nu}$, where the atoms group in pairs, with zero populations for odd particle numbers, which according to Eq.~\ref{eq:phaseandcat} is an indication that $\rho_{th}$ is close to the cat state $\ket{cat(z=0)}$. This is confirmed by Fig.~\ref{fig:Fig.2}(a2) (red trace), which shows the projection $Tr( \rho_{th} \ket{cat(z)}\bra{cat(z)})$ onto the cat state $\ket{cat(z)}$. The plot shows that a fidelity of nearly unity is reached for $z=0$. Accordingly, the projection $Tr( \rho_{th} \ket{\pi z}\bra{\pi z}$ onto the phase state $\ket{\pi z}$ in the blue trace of Fig.~\ref{fig:Fig.2}(a2) shows two peaks at $\pm \pi/2$, where values near 0.5 are attained. For odd particle numbers $N$, a pure ground state is not to be expected even at zero temperature, since the lowest energy eigenstate then exhibits perfect twofold degeneracy. Even for low values of $N$ at least a 2D manifold of states with equal energies contribute to $\rho_{th}$, and hence the ground state incured by the system is typically a mixed state for arbitrarily low temperatures.

\begin{figure}
\includegraphics[scale=0.7, angle=0, origin=c]{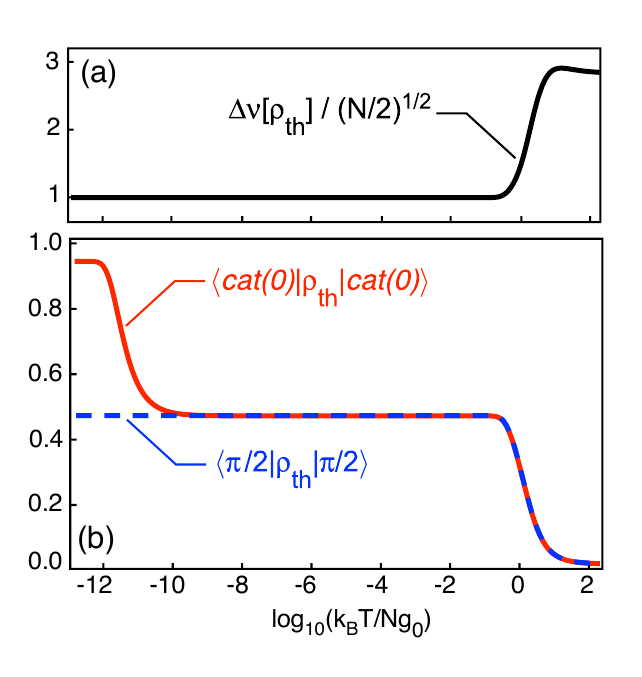}
\caption{The variance $\Delta \nu / \sqrt{(N/2)}$ of $\bra{\nu}\rho_{th}\ket{\nu}$ (a), $\bra{cat(0)}\rho_{th}\ket{cat(0)}$ (red solid trace in(b)) and $\bra{\pi /2}\rho_{th}\ket{\pi /2}$ (blue dashed trace in(b))  are plotted versus $\log_{10}(k_B T/ N g_0)$ over 15 decades.}
\label{fig:Fig.3}
\end{figure}

Returning to even $N$, for higher but still extremely low temperatures ($k_B T = 3 \times10^{-12} Ng_0$) one can see how the cat state vanishes in favour of a mixed state. In Fig.~\ref{fig:Fig.2}(b) the staggered part of the populations is decreased, while the projection onto the cat state $\ket{cat(z)}$ assumes a maximal value 0.65 at $z=0$, although the overlap with the phases states $\ket{\pm \pi /2}$ remains the same. For even larger temperatures in the range $10^{-11} < k_B T / N g_0 < 10^{-1}$ one finds practically the same results over 10 decades shown in Fig.~\ref{fig:Fig.2}(c). In this regime the width of the population distribution (c1) is determined by quantum fluctuations, the projection onto $\ket{cat(z)}$ (c2) is practically 1/2 for any value of $z$, while the projections onto $\ket{\pm \pi /2}$ (c2) remain as in (a2) and (b2). This shows that $\rho_{th} \approx \frac{1}{2}(\ket{\pi/2}\bra{\pi/2)}+\ket{-\pi/2}\bra{-\pi/2)})$ in good approximation is given by the incoherent mixture of the phase states $\ket{\pm \pi/2}$. As the temperature is further increased by just a factor 10 ($k_B T = Ng_0$), thermal noise becomes dominant and the situation changes again according to Fig.~\ref{fig:Fig.2}(d). Now the population distribution (d1) notably broadens and the projections onto the states $\ket{cat(z)}$ and  $\ket{\pm \pi /2}$ notably decrease. In all plots of the population distributions the values of $Tr(L^2 \rho_{th})$ are indicated. Even for the largest temperature shown, where the overlap with the phase states $\ket{\pm \pi /2}$ is significantly reduced, the angular momentum per particle, attaining the value 0.866, remains close to unity. Although the thermal state in the case of Fig.~\ref{fig:Fig.2}(d) has a notably broader distribution of populations in the Fock basis (d1) as compared to that in (c1), in terms of its phase properties it is still similar to an incoherent superposition of the phase states $\ket{\pm \pi /2}$. 

The temperature dependence of the thermal ground state is summarized in Fig.~\ref{fig:Fig.3}, where the standard deviation $\Delta \nu$ of $\bra{\nu}\rho_{th}\ket{\nu}$ as well as $\bra{cat(0)}\rho_{th}\ket{cat(0)}$ and $\bra{\pi /2}\rho_{th}\ket{\pi /2}$ are plotted versus $\log_{10}(T/ N g_0)$ over 15 decades. In (a) between $-12$ and $-1$ one sees constant $\Delta \nu = \sqrt{N/2}$, which represents Poissonian quantum noise for $N/2$ particles in each mode. Above $\log_{10}(T/ N g_0) = -1$ thermal noise begins to dominate such that $\Delta \nu$ rapidly grows. In (b) it is seen that between $-13$ and $-12$ the system state has nearly unity overlap with the cat state $\ket{cat(0)}$. Between $-12$ and $-11$ the cat state rapidly decays such that between $-11$ and $-1$ the system state is well described by an incoherent superposition of $\ket{\pm \pi /2}$. Above $-1$, the phase states $\ket{\pm \pi /2}$ begin to decohere and the systems becomes thermal. This decoherence of the phase states is also seen in $\bra{\pi /2}\rho_{th}\ket{\pi /2}$ in (c). A central message behind these observations is that with increasing $N$, the ground state rapidly acquires a two-fold degeneracy, and the system state even for extremely low or, if $N$ is odd, even for zero temperature becomes an incoherent mixture of phase states. A mean field description of this scenario would require the concept of spontaneous symmetry breaking. Each of these phase states itself is a maximally coherent (quasi classical) state, which only decoheres at a many orders of magnitudes higher temperature than that required for splitting up the cat state for even $N$ into an incoherent superposition of phase states. Up to relatively high temperatures on the order of $k_B T = N g_0$, the ground state is an incoherent mixture of two states with orbital angular momenta close to $\pm 1$ per particle. Note also the discussions for zero temperature in Refs.~\cite{Bad:09,  Fis:13, Zhu:15}.

\begin{figure}
\includegraphics[scale=0.35, angle=0, origin=c]{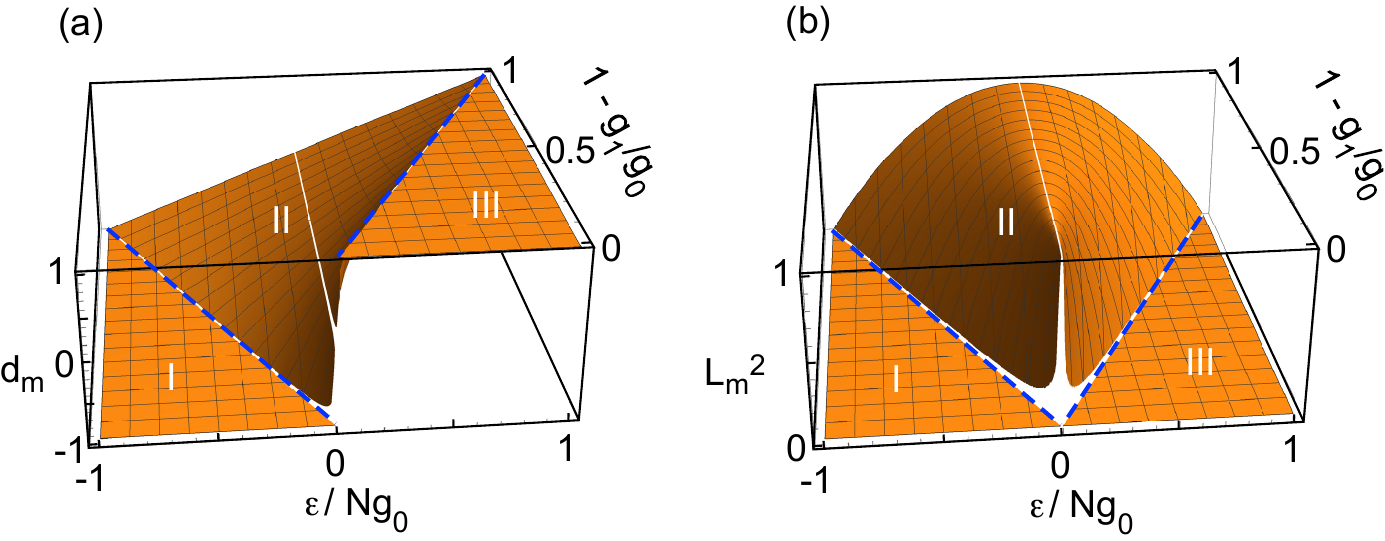}
\caption{(a) Mean-field calculation of the difference of the populations of the single-particle modes $d_m$ as a function of $\varepsilon/ Ng_0$ and $1- g_1/g_0$. The dashed blue lines indicate $2^{\textrm{nd}}$ order phase boundaries, where time-reversal symmetry is spontaneously broken. (b) Corresponding square of the mean angular momentum.}
\label{fig:Fig.4}
\end{figure}

\section{Ground state phase diagram}
\subsection{Zero temperature}
The quantities considered in this section, in order to characterize different phases of the system, are the expectation value $d \equiv \frac{1}{N} \langle D \rangle$ and the fluctuations $\Delta d \equiv  \frac{1}{N}\Delta D$, where $D \equiv a_{1}^{\dagger} a_{1 }- a_{2}^{\dagger} a_{2}$ denotes the population difference of the single-particle modes and $\Delta D \equiv \langle (D-\langle D \rangle)^{2} \rangle^{1/2}$. For the ground state, these quantities are calculated as functions of $\varepsilon$ and the collision parameters $g_0$ and $g_1$, and the full quantum result for $d$ is compared to a mean field calculation. 

Let us begin with the mean field calculation by replacing the matter field operator $\psi$ in the general Hamiltonian at the beginning of Sec.II by a complex wave function, which is written as the most general superposition of the two basis modes $\psi = \sqrt{N} (\cos(\theta) \alpha_{1} + \sin(\theta) e^{i \phi}\alpha_{2})$. For this scenario, the mean field value of $d$ is given by $d_{m} = \frac{1}{N}(|\langle \alpha_1 \ket{\psi}|^2 - |\langle \alpha_2 \ket{\psi}|^2) = \cos(2\theta)$. Setting $\rho_1=\rho_2$ and $\rho_3=\rho_4 =0$ as in the context of Eq.~\ref{eq:Hamiltonian}, one obtains
\begin{eqnarray}
\label{eq:meanfield}
H[\theta, \phi] &=& N \frac{1}{2} (\sin^2(\theta)-\cos^2(\theta))\, \varepsilon
\\ \nonumber
&+& \frac{1}{2} g \rho_0 N^2 (\cos^4(\theta)+\sin^4(\theta)) 
\\ \nonumber
&+& g \rho_1 N^2 (1+2 \cos^2(\phi)) \sin^2(\theta) \cos^2(\theta)
\end{eqnarray}
Our task here is to find the minimum of $H[\theta, \phi]$ with respect to $\theta$ and $\phi$. Only the last term depends on $\phi$ and is obviously minimized by setting $\phi = \pm \pi/2$ for arbitrary values of $\theta$ and a repulsive collision parameter $g>0$. Hence, $H[\theta,\pm\pi/2] = - N \frac{1}{2} \varepsilon \cos(2\theta) + \frac{1}{4} (g_1 - g_0) N^2 \sin^2(\theta) + \frac{1}{4} N^2 g_0$, which is minimized with respect to $\theta$ if $\varepsilon = \cos(2\theta) (g_0-g_1) N$. Noting that $\cos(2\theta)$ is necessarily constrained to the interval $[-1,1]$, the simple relation
\begin{eqnarray}
\label{eq:meanfield}
d_m &=& \frac{\varepsilon}{(g_0-g_1) N} \, \chi_{[-1,1]}\left(\frac{\varepsilon}{(g_0-g_1) N}\right)  \\ \nonumber
&+& \left[1 - \chi_{[-1,1]}\left(\frac{\varepsilon}{(g_0-g_1) N}\right)\right] \, \textrm{Sign}\left(\frac{\varepsilon}{(g_0-g_1) N}\right) ,
\end{eqnarray}
is obtained, where $\chi_{[-1,1]}$ denotes the indicator function for the interval $[-1,1]$. The mean field result $d_m$ is plotted as a function of $\varepsilon/ Ng_0$ and $1- g_1/g_0$ in Fig.~\ref{fig:Fig.4}(a). Three distinct regions are identified (denoted I, II, III). Where $d_m$ takes the values $-1$ or $+1$, only one of the modes $\alpha_i$ is populated, while in between these regions both modes are superimposed with a relative phase randomly taking one of the values $\pm \pi/2$. The three regions are separated by phase boundaries, indicated by dashed blue lines, where time-reversal symmetry is spontaneously broken. Note that $d_m$ can be expressed in terms of the mean field value $L_{m}$ of the previously defined angular momentum. With $a_1$ and $a_2$ replaced by their mean values $\sqrt{N}\cos(\theta)$ and $i \sqrt{N}\sin(\theta)$, respectively, one gets $L_{m} = 2\sin(\theta)\cos(\theta) = \sin(2\theta)$ and thus $L_{m}^2 = 1-d_m^2$. Similarly as $d_m$, this quantity, which is plotted in Fig.~\ref{fig:Fig.4}(b), may serve as an order parameter discriminating the chiral phase II from the time-reversal symmetric phases I and III. Analogous mean field results have been published for a specific implementation of the general model considered here, where the two single-particle modes are chosen to be Bloch functions of two degenerate high-symmetry points in the second band of a 2D optical lattice \cite{Isa:05, Liu:06, Cai:11, Oel:13, Li:16, Koc:16}. In the context of this example, which will be discussed in more detail in Sec.VII, an intuitive explanation has been given, why a superposition of the single-particle modes with a relative phase $\pm \pi/2$ is energetically favoured. The reason is that this superposition allows the atoms to optimally avoid each other thus minimizing their repulsive interaction  \cite{Isa:05, Liu:06}. More recently, mean field phase diagrams for double well scenarios in configuration space with pair-tunneling have been discussed in Refs.~\cite{Rub:17, Agb:18}.

\begin{figure}
\includegraphics[scale=0.35, angle=0, origin=c]{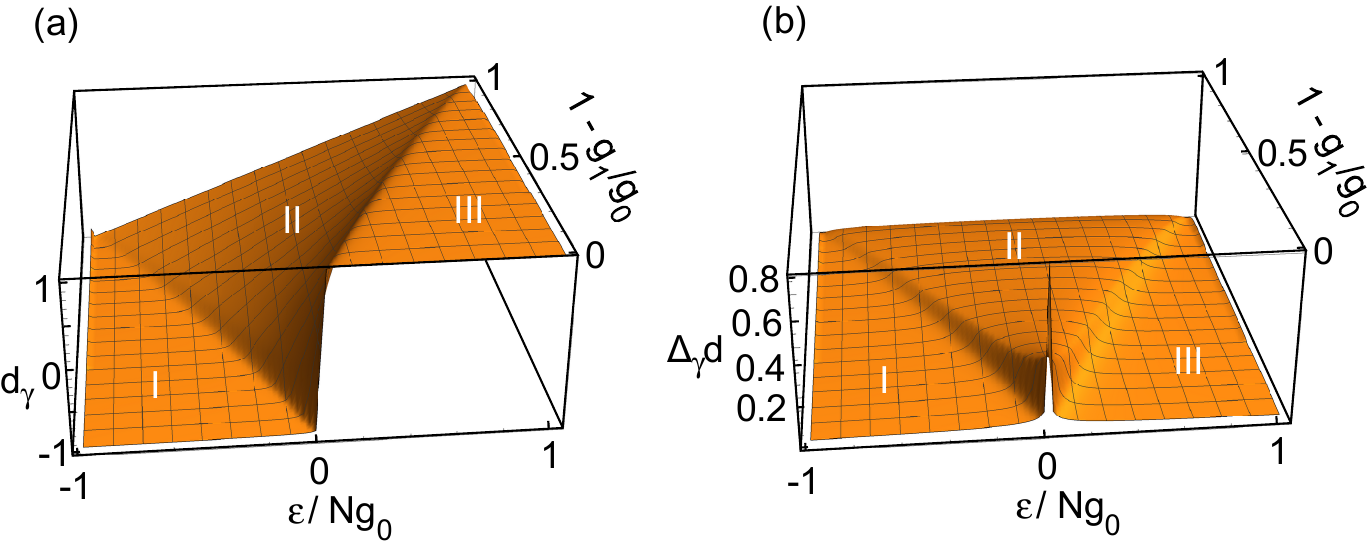}
\caption{Full quantum calculation of $d_{\gamma}$ (a) and $\Delta_{\gamma} d$ (b) plotted against $\varepsilon/Ng_0$ and $1- g_0/g_1$ for $N=100$.}
\label{fig:Fig.5}
\end{figure}

Next, restricting ourselves to the case of even $N$, the expectation value $d_{\gamma} \equiv \frac{1}{N} \langle D \rangle_{\gamma}$ and the standard deviation $\Delta_{\gamma} d \equiv  \frac{1}{N}\Delta_{\gamma} D$ are determined for the non-degenerate zero temperature ground state $\ket{\gamma} = \sum_{\nu=0}^{N} \gamma_{\nu} \ket{N-\nu,\nu}$ of the Hamiltonian $H$ in Eq.~\ref{eq:Hamiltonian}, which leads to
\begin{eqnarray}
\label{eq:phasediagram_gnd}
d_{\gamma} &=& 1 - \frac{2}{N} \, \bar{\nu}_{\gamma},\,\, \bar{\nu}_{\gamma} \equiv \sum_{\nu=0}^{N} |\gamma_{\nu}|^2 \nu \, ,
\\ \Delta_{\gamma} d &=& \frac{2}{N} \sqrt{\sum_{\nu=0}^{N} |\gamma_{\nu}|^2 (\nu-\bar{\nu}_{\gamma})^2}
\end{eqnarray}
The amplitudes $\gamma_{\nu}$ are obtained by direct diagonalisation of $H$ in the Fock basis. In Fig.~\ref{fig:Fig.5}, $d_{\gamma}$ (a) and $\Delta_{\gamma} d$ (b) are plotted against $\varepsilon/Ng_0$ and $1- g_0/g_1$ for $N=100$. The graph in (a) shows good agreement with the mean-field results of Fig.~\ref{fig:Fig.4}. Only at the phase boundaries (dashed lines in Fig.~\ref{fig:Fig.4}) the finite size of $N$ in the full quantum description smoothes out the kinks seen in the mean-field diagram. As $N$ approaches infinity, the kinks of the mean-field diagram are reproduced. The fluctuations plotted in Fig.~\ref{fig:Fig.5} (b) reproduce the structure of the phase diagram in (a). In regions I and III, nearly all particles populate a single mode with very small fluctuations. In the central region II, both modes are populated with the result of notable fluctuations. At the point $\varepsilon/Ng_0 = 1- g_1/g_0 =0$ there is a pronounced peak of the fluctuations with $\Delta_{\gamma} d(0,0) = 1/\sqrt{2}$, i.e., $\Delta_{\gamma} D$ scales with $N$ showing strongly super-Poissonian behaviour. At this point $g_1=g_0$, i.e., the total interaction in Eq.~\ref{eq:Hamiltonian} becomes $\frac{1}{2}\,g_0\,N^2 (1 - L^2)$. Hence, since $L^2 \approx 1$ in region II, the total interaction vanishes. This resembles the behaviour of a non-interacting BEC at the critical temperature. As $1- g_1/g_0$ grows, the fluctuations decrease, approaching zero for $1- g_1/g_0 =1$. It is interesting to note that for $1- g_1 / g_0 =\frac{2}{3}$ one finds $\Delta_{\gamma} d = 1/ \sqrt{N}$, i.e. Poissonian fluctuations. This case, for example, naturally occurs in a specific implementation of the general model by using the $p_x$ and $p_y$-orbitals of a 2D harmonic oscillator as single-particle modes.

\begin{figure}
\includegraphics[scale=0.38, angle=0, origin=c]{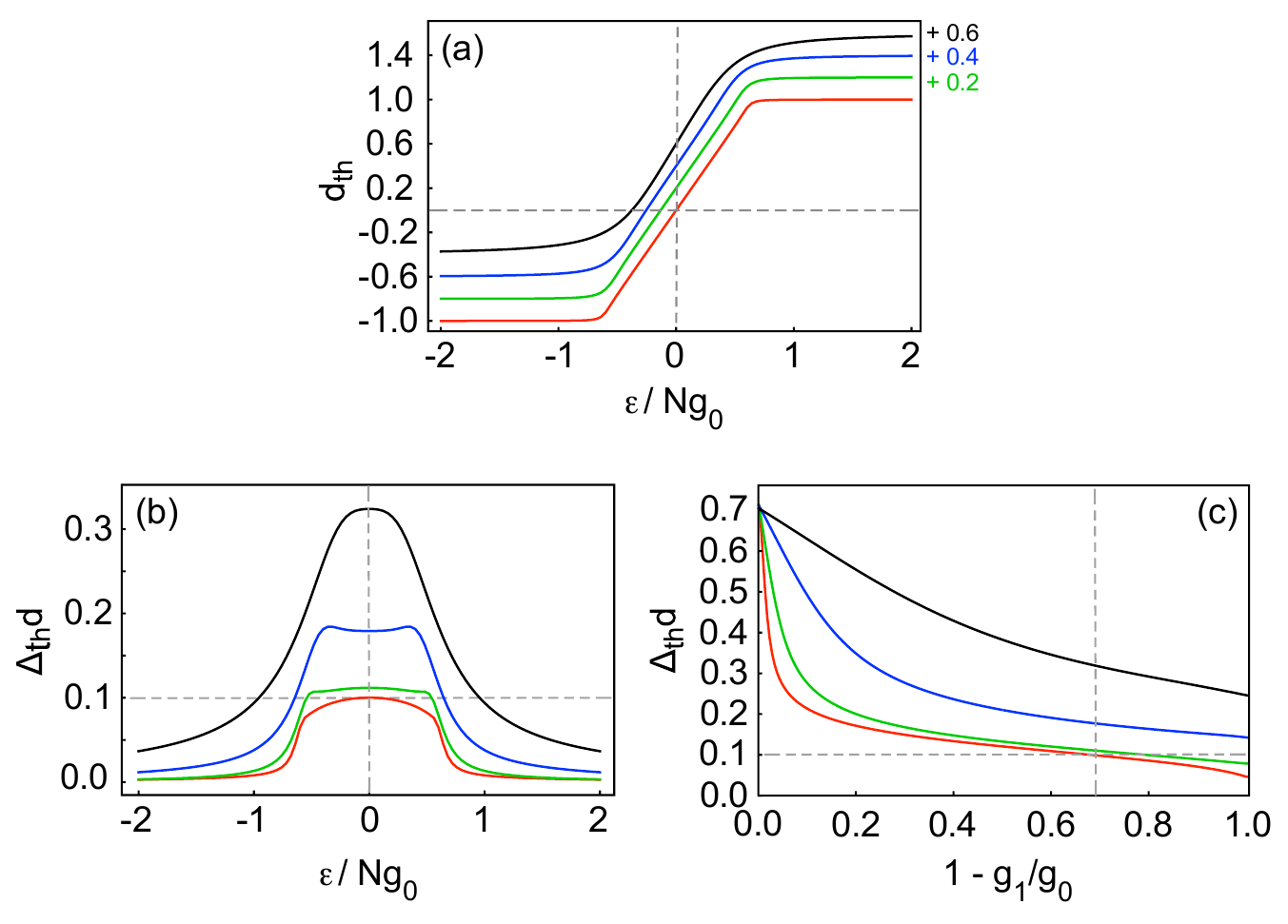}
\caption{Quantum calculations of $d_{\textrm{th}}$ (a) and $\Delta_{\textrm{th}} d$ (b) for $1- g_1/g_0 = 2/3$ and $N=100$ plotted against $\varepsilon/N g_0$. 
In (c) $\Delta_{\textrm{th}} d$ is plotted versus $1 - g_1/g_0$ for $\varepsilon = 0$. In each panel, graphs are shown for four temperatures $k_B T/ N g_0 \in\{0.1, 0.3, 1, 3\}$ in ascending order. In (a), for optimal visibility, the uppermost three graphs are shifted upwards vertically by values $0.2, 0.4, 0.6$. The horizontal dashed gray lines in (b) and (c) mark Poissonian noise for $N=100$.}
\label{fig:Fig.6}
\end{figure}

\subsection{Finite temperature}
The phase diagram in Fig.~\ref{fig:Fig.5} is readily extended to account for finite temperatures by calculating 
\begin{eqnarray}
\label{eq:phasediagram_th}
d_{\textrm{th}} &=& 1 - \frac{2}{N} \, \bar{\nu}_{\textrm{th}},\,\, \bar{\nu}_{\textrm{th}}\equiv \sum_{\nu=0}^{N} \bra{\nu}\rho_{th}\ket{\nu} \nu \, ,
\\ \Delta_{\textrm{th}} d &=& \frac{2}{N} \sqrt{\sum_{\nu=0}^{N} \bra{\nu}\rho_{th}\ket{\nu} (\nu-\bar{\nu}_{\textrm{th}})^2}
\end{eqnarray}
with $\rho_{th}$ according to Eq.~\ref{eq:thermalstate}. In Fig.~\ref{fig:Fig.6}(a) and (b), sections through the plots in Fig.~\ref{fig:Fig.5}(a) and (b) are shown at $1- g_1/g_0 = 2/3$ for four different temperatures $k_B T/ Ng_0 \in\{0.1, 0.3, 1, 3\}$. Note in (a) that the kinks recognized in the lowermost (red) graph with the lowest shown temperature $k_B T/ Ng_0 = 0.1$ soften as the temperature is increased. As seen in (b), at $k_B T/ Ng_0 = 0.1$, quantum fluctuations dominate, which for $\varepsilon = 0$ reach a maximum $\Delta_{\textrm{th}} d = 1/\sqrt{N}$ (i.e., $0.1$ for $N=100$), which corresponds to Poissonian noise. In Fig.~\ref{fig:Fig.6}(c) the fluctuations for $\varepsilon = 0$ are plotted versus $1- g_1/g_0$, showing the pronounced maximum at $1- g_1/g_0 = 0$, which was already found for the zero-temperature ground state in Fig.~\ref{fig:Fig.5}(b). Here, regardless of the temperature, $\Delta_{\textrm{th}} d = 1/\sqrt{2}$ and hence $\Delta_{\textrm{th}} D$ scales linearly with $N$, i.e., the fluctuations acquire the same strongly super-Poissonian character for all temperatures.

\begin{figure}
\includegraphics[scale=0.26, angle=0, origin=c]{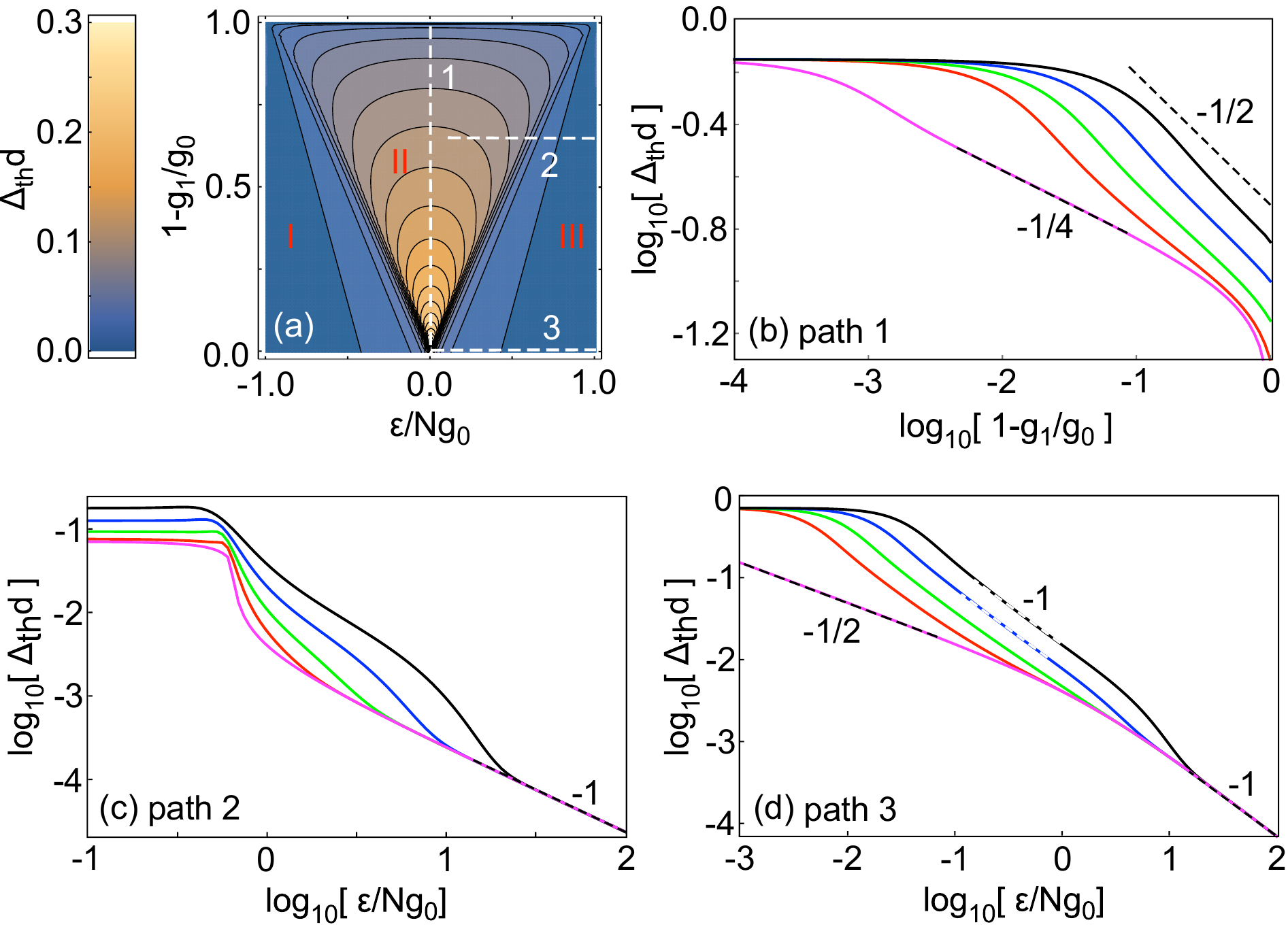}
\caption{(a) repeats the phase diagram in Fig.~\ref{fig:Fig.5} as a contour-plot with the three phases I , II, and III, in order to illustrate three paths labeled 1,2,3 that are examined in (b), (c) and (d). Dashed lines in (b), (c) and (d) emphasize power law behaviour labeled with the corresponding critical exponent. In all plots $N=200$.}
\label{fig:Fig.7}
\end{figure}
In order to characterize the nature of the phase boundaries between phases I, II and III and the critical point at $(\varepsilon,1- g_1/g_0)=(0,0)$, double-log plots of $\Delta_{\textrm{th}} d$ are shown in Fig.~\ref{fig:Fig.7} for different temperatures along three different paths (indicated 1,2,3 in the phase diagram in (a)). In (b) the critical point at the origin is approached from the chiral phase II along the vertical $1- g_1/g_0$ axis (path 1). The five shown graphs are for the temperatures $k_B T/ Ng_0 \in\{0, 0.25, 0.5, 1, 2\}$ in ascending order. The black dashed lines indicate regions of power law behaviour with critical exponents -1/4 for the zero temperature case and -1/2 approximately describing the finite temperature cases. Analogous plots are shown in (c) and (d) for the paths indicated 2 and 3 in (a). In these plots the phase boundary between the phases II and III is explored at the two values $1- g_1/g_0  = 2/3$ (c) and $1- g_1/g_0  = 0$ (d). Again, the dashed lines indicate regions of power law behaviour labeled with the corresponding critical exponents. For example, in (d), where the critical point at the origin is approached from phase III along the $\varepsilon$-axis, the critical exponent for the zero-temperature case switches from -1 to -1/2. 

\subsection{Pair correlations}
Further insight into the nature of the phases I, II ,III is obtained by determining the condensate fraction in terms of the Penrose-Onsager criterion \cite{Pen:56} by considering the eigenvalues $\lambda_{\pm}^{s}$ of the single-particle density matrix (SPDM) $\langle a_{n}^{\dagger}a_{m}\rangle$ with $n,m \in \{1,2\}$. The fragmentation associated with the SPDM, defined in accordance with Refs.~\cite{Bad:09, Fis:13, Zhu:15} as $F_{s} \equiv 1- \frac{|\lambda_{+}^{s} -\lambda_{-}^{s}|}{\langle n_{1}\rangle+\langle n_{2}\rangle}$, takes the general form 
\begin{eqnarray}
\label{eq:Fragment1}
F_{s} =  1 - \frac{\sqrt{(\langle n_{1}\rangle-\langle n_{2}\rangle)^2 + 4 |\langle a_{1}^{\dagger}a_{2}\rangle|^2}}{\langle n_{1}\rangle+\langle n_{2}\rangle} \, .  
\end{eqnarray}
For an arbitrary state given by a density operator $\rho$, we can straight forwardly evaluate the quantities
\begin{eqnarray}
\label{eq:Fragment2}
\nonumber
\langle a_{1}^{\dagger} a_{2} \rangle \,\,\, &=&  \sum_{\nu=0}^{N} \, \sqrt{(N-(\nu-1)) \,\nu } \,\, \langle \nu | \rho | \nu -1 \rangle 
\\  \\ \nonumber
\langle n_{1} - n_{2}\rangle &=&  \sum_{\nu=0}^{N} (N-2\nu) \, \langle \nu | \rho | \nu \rangle
\end{eqnarray}
in order to numerically determine $F_{s}$. For density operators $\rho =  \sum_{\nu=0}^{N} \rho^{{(n)}}\ket{E_n} \bra{E_n}$, diagonal in the eigenbasis $\ket{E_n}$ (e.g., thermal states), one finds $\langle a_{1}^{\dagger} a_{2} \rangle = 0$ and hence $F_{s} =  1 - \frac{ |\langle n_{1}\rangle-\langle n_{2}\rangle|}{\langle n_{1}\rangle+\langle n_{2}\rangle}$. The reason is that the eigenstates $\ket{E_n}$ are superpositions of  number states $\ket{\nu}$ with either even or odd values of $\nu$ (cf. Sec.III).

These considerations can be readily extended to the pair density matrix (PDM) $\langle b_{n}^{\dagger}b_{m}\rangle$ with $n,m \in \{1,2\}$, where $b_n \equiv a_n a_n$ denotes the pair anihilation operator and $p_n \equiv b_{n}^{\dagger}b_{n}$ the pair number operator. The corresponding fragmentation $F_{p} \equiv 1- \frac{|\lambda_{+}^{p} -\lambda_{-}^{p}|}{\langle p_{1}\rangle+\langle p_{2}\rangle}$, associated with the eigenvalues $\lambda_{\pm}^{p}$ of the PDM, reads
\begin{eqnarray}
\label{eq:Fragment3}
F_{p} =  1 - \frac{\sqrt{(\langle p_{1}\rangle-\langle p_{2}\rangle)^2 + 4 |\langle b_{1}^{\dagger}b_{2}\rangle|^2}}{\langle p_{1}\rangle+\langle p_{2}\rangle} \, ,  
\end{eqnarray}
and
\begin{eqnarray}
\label{eq:Fragment2}
\nonumber
\langle b_{1}^{\dagger}b_{2}\rangle \, \, \, \,  &=& \sum_{\nu=0}^{N} \sqrt{(N-(\nu-1)) (N-(\nu-2))\,\nu(\nu -1)} \,
\\ &\times& \quad \langle \nu | \rho | \nu -2 \rangle 
\\ \nonumber
\langle p_{1} - p_{2}\rangle &=& \sum_{\nu=0}^{N} (N^2 - N (2\nu+1) + 2 \nu) \, \langle \nu | \rho | \nu \rangle
\\ \nonumber
\langle p_{1} + p_{2}\rangle &=& \sum_{\nu=0}^{N} (N^2 - N (2\nu+1) + 2 \nu^2) \, \langle \nu | \rho | \nu \rangle \, .
\end{eqnarray}
In Fig.~\ref{fig:Fig.8} the single-particle fragmentation $F_{s}$ (a) and the pair fragmentation $F_{p}$ (b) are plotted versus the chemical potential difference $\varepsilon / N g_0$ of the single particle modes with $g_1/g_0 = 1/2$, thus intersecting all three phases I, II, and III. A thermal state as in 
Eq.~\ref{eq:thermalstate} is assumed with increasing temperatures $k_B T/ N g_0 \in\{0.1, 0.3, 1, 3\}$ represented by the colors red, green, blue and black. For the phases I and III, for all temperatures shown, $F_{s}$ is notably smaller than 0.5, associated with the fact that most of the atoms populate the same single-particle mode $\alpha_i$ where they form a condensate. Towards the center of phase II (around $\varepsilon = 0$), the single-particle fragmentation rises to unity. This indicates that the realized state can no longer be described as a single multiply populated quasi-particle state, i.e. a condensate. In fact, as discussed in Sec.IV, for the shown temperatures, in the vicinity of $\varepsilon = 0$, the realized state is an incoherent superposition of two condensates approximately described by the phase states $\ket{\pm \pi/2}$. However, as seen in (b), the pair fragmentation $F_{p}$ for sufficiently low temperature remains close to zero everywhere, such that, in terms of pairs, phase II maintains the character of a condensate.

\subsection{Angular momentum and entanglement entropy}

\begin{figure}
\includegraphics[scale=0.36, angle=0, origin=c]{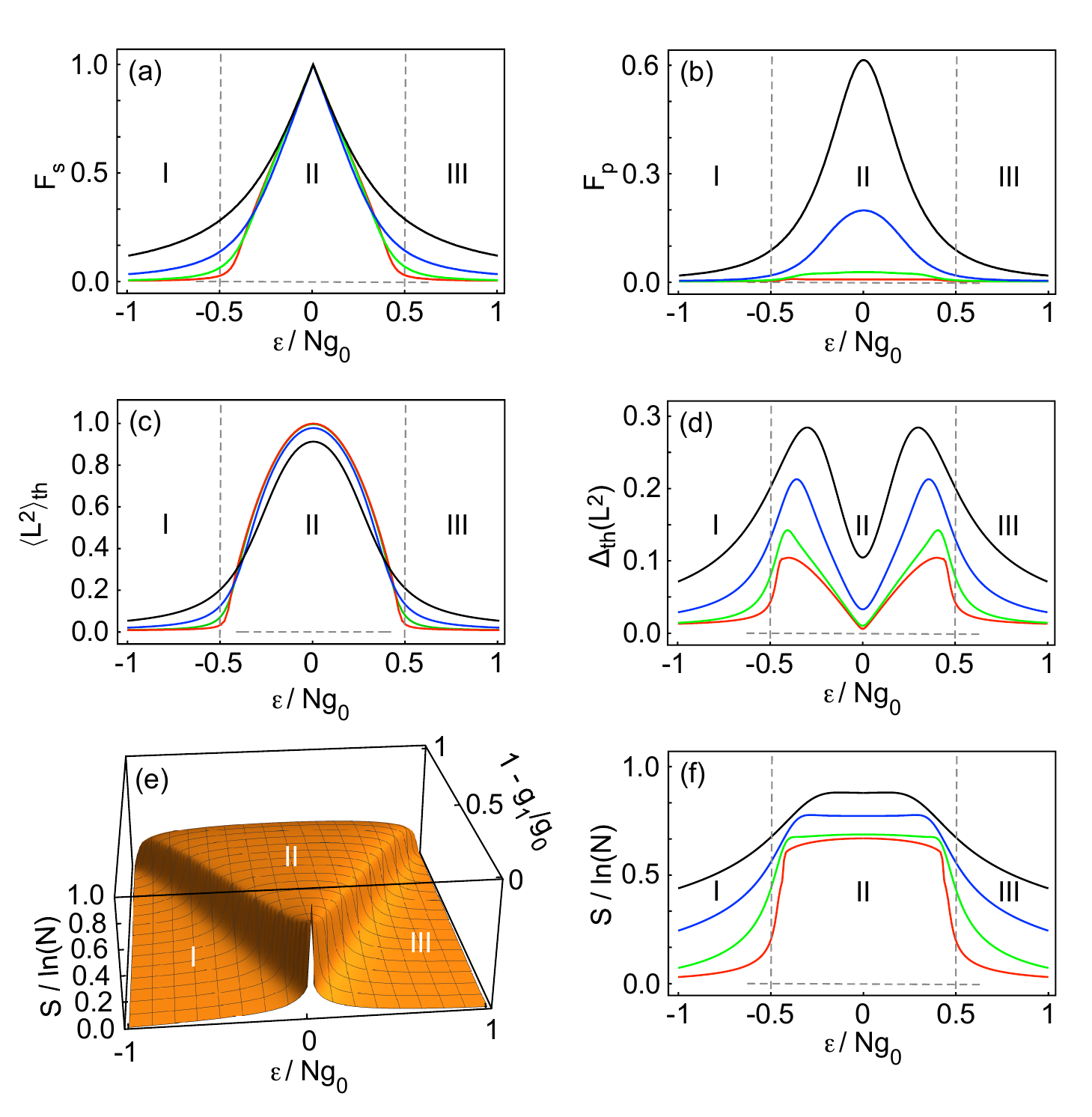}
\caption{In (a) and (b), the single-particle fragmentation $F_{s}$ and the pair fragmentation $F_{p}$ are plotted versus $\varepsilon / N g_0$, respectively, for fixed $1- g_1/g_0 = 0.5$ and four temperatures $k_B T/ N g_0 \in\{0.1, 0.3, 1, 3\}$ (red, green, blue, black). In (c) and (d), the expectation value $\langle L^2 \rangle_{\textrm{th}}$ (c) and the associated fluctuations $\Delta_{\textrm{th}}(L^2)$ (d) are plotted versus $\varepsilon / N g_0$ for fixed $1- g_1/g_0 = 0.5$ and four temperatures $k_B T/ N g_0 \in\{0, 0.3, 1, 3\}$ in ascending order (dashed red, green, blue, black). (e) The entanglement entropy $S$ (cf. Eq.~\ref{eq:EtEntropy}) of the zero temperature groundstate is plotted in units of $\ln(N)$ versus $\varepsilon$ and $1- g_1/g_0$. (f) The entanglement entropy $S$ is shown for thermal states with temperatures $k_B T/ Ng_0 \in\{0, 0.3, 1, 3\}$ and fixed $1- g_1/g_0 = 0.5$. In (a) and (b) $N=50$, for all other graphs $N=200$.}
\label{fig:Fig.8}
\end{figure}

It is interesting to characterize the possible phases in terms of their angular momentum using a full quantum description. The expectation value of $L$ vanishes for states parametrized by density operators diagonal with respect to the eigenbasis $\ket{E_n}$ (e.g., for thermal states) since $\langle a_{1}^{\dagger} a_{2} \rangle = 0$ for such states, as discussed above. For symmetry reasons, in a full quantum description, macroscopic angular momentum should indeed not occur. Nevertheless, $ L^2$ can have a non-zero expectation value. Since the flavour changing interaction, which is proportional to $L^2$ (cf. Eq.~\ref{eq:Hamiltonian2}), can be viewed as a pair tunneling process, $\langle L^2 \rangle$ can be interpreted as an order parameter indicating the presence of coherent pairs. In Fig.~\ref{fig:Fig.8}, $\langle L^2 \rangle_{\textrm{th}}$ (c) and the associated standard deviation $\Delta_{\textrm{th}}(L^2) = \sqrt{\langle L^4 \rangle_{\textrm{th}} -\langle L^2 \rangle_{\textrm{th}}^2}$ (d) are plotted as $\varepsilon$ is tuned across the phase boundaries from phase I to III for $1- g_1/g_0 = 0.5$. The shown graphs are for thermal states with temperatures $k_B T/ Ng_0 \in\{0, 0.3, 1, 3\}$. At $\varepsilon = 0$, $\langle L^2 \rangle_{\textrm{th}}$ is maximized, in accordance to the mean field result in Fig.~\ref{fig:Fig.4}(d), becoming unity for the case of zero temperature. Hence, the fluctuations of $L^2$ must attain a minimum at $\varepsilon = 0$, which for low temperatures rapidly approaches zero as $N$ is increased, allthough according to Fig.~\ref{fig:Fig.6}(b) the fluctuations $\Delta_{\textrm{th}}d$ of the relative population difference $d_{\textrm{th}}$ take a maximum.

Another instructive quantity is the entanglement entropy of the ground state at zero or finite temperature with respect to the sub-spaces associated with each of the single-particle modes $\alpha_i, i\in\{1,2\}$. This quantity determines the increase of ones ignorance due to bipartite entanglement if one of the the single-particle modes is traced out. For the thermal state in Eq.~\ref{eq:thermalstate}, the general expression for the entanglement entropy is $S \equiv - Tr^{(1)}[\rho_{th}^{(1)} \ln [\rho_{th}^{(1)}]]$, with $\rho_{th}^{(1)} \equiv Tr^{(2)}[\rho_{th}]$ and $Tr^{(i)}$ denoting the trace with respect to the sub-systems associated with modes $\alpha_i$. One readily obtains
\begin{eqnarray}
\label{eq:EtEntropy}
S = - \sum_{\nu=0}^{N} \bra{\nu}\rho_{th}\ket{\nu} \ln(\bra{\nu}\rho_{th}\ket{\nu}) \, ,  
\end{eqnarray}
using the populations $\bra{\nu}\rho_{th}\ket{\nu}$ plotted in Fig.~\ref{fig:Fig.2}. In Fig.~\ref{fig:Fig.8}(e), $S$ is plotted in units of its maximally possible value $\ln(N)$ versus $\varepsilon$ and $1- g_1/g_0$ for zero temperature. In Fig.~\ref{fig:Fig.8}(f), graphs for thermal states with temperatures $k_B T/ Ng_0 \in\{0, 0.3, 1, 3\}$ are shown for $1- g_1/g_0 = 0.5$. For zero temperature, i.e., when the ground state is practically a pure state that has zero entropy, the large entanglement entropy, seen in the chiral phase (region II), is completely due to the presence of massive entanglement between the single-particle mode sub-spaces. For larger temperatures, a large part of the entanglement entropy reflects the non-zero entropy of the thermal state $\rho_{th}$.

\begin{figure*}
\includegraphics[scale=0.98, angle=0, origin=c]{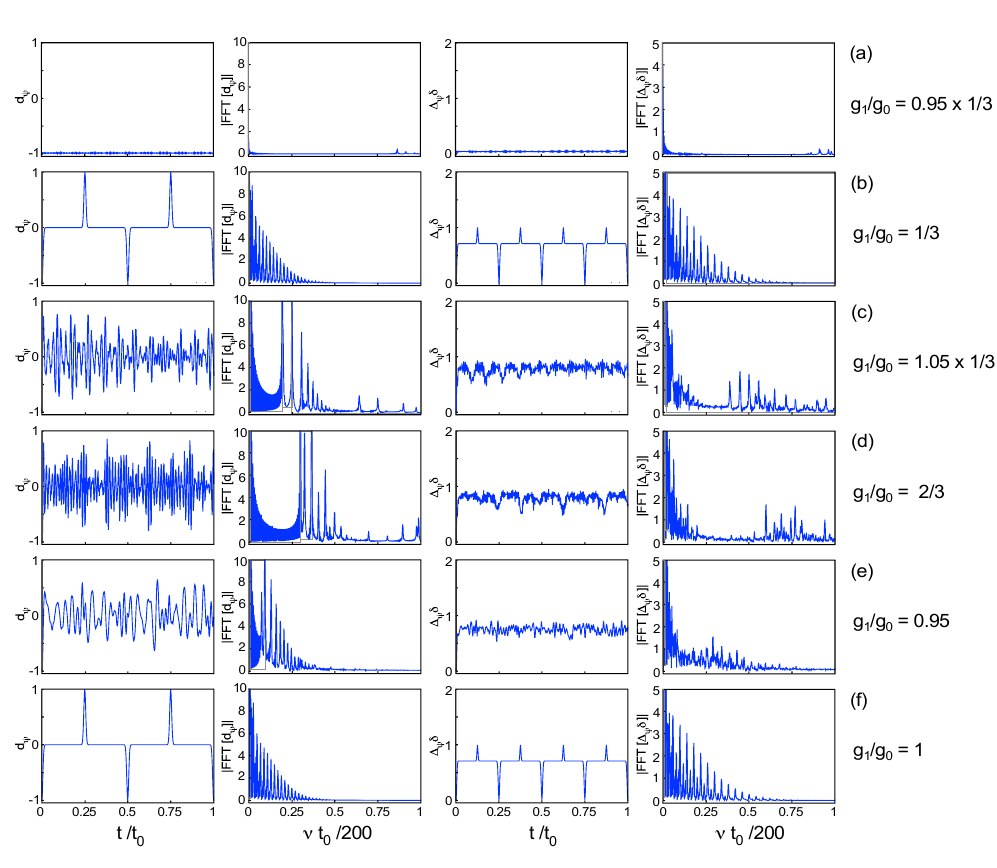}
\caption{(color online). Self-trapping and pair tunneling dynamics for the case of degenerate modes ($\varepsilon = 0$). The rows labeled (a-f) correspond to the choices of $g_1$ indicated. Each row from left to right shows $d_{\psi(t)}$, $\mathcal{F}[d_{\psi(t)}]$, $\Delta_{\psi(t)}d$, and $\mathcal{F}[\Delta_{\psi(t)}d]$ with $\mathcal{F}[x(t)]$ denoting the Fourier spectrum of $x(t)$. The time and frequency axes are scaled according to $\tilde t = t/t_0$ and $\tilde \nu =  \nu t_0/200$, respectively, with $t_0 \equiv \frac{2\pi \hbar}{g_{1}}$.}
\label{fig:Fig.9}
\end{figure*}

\section{Self-trapping}
The Hamiltonian of Eq.~\ref{eq:Hamiltonian} shows rich non-linear dynamics. A notable phenomenon also found in the two-site Hubbard model is self-trapping \cite{Sme:97, Mil:97, Rap:12, Alb:05, Lev:07}, i.e., an interaction induced suppression of tunneling, where for our model tunneling refers to pair-tunneling resulting via flavour changing interaction. An initial state $\ket{\psi(0)} = \ket{N,0}$ at $t=0$ is assumed with all $N$ atoms piled up in mode $\alpha_1$. For later times
\begin{eqnarray}
\label{eq:dynamics}
\ket{\psi(t)} = \sum_{n=0}^{N} \, \ket{E_{n}}  \bra{E_{n}}\psi(0)\rangle  \, e^{-\frac{i}{\hbar} E_{n} t} \,\, ,
\end{eqnarray}
is determined and the time evolution and the associated Fourier spectra of the expectation value $d_{\psi(t)} = \frac{1}{N}\bra{\psi(t)} D\,|\psi(t)\rangle $ and the corresponding fluctuations $\Delta_{\psi(t)}d$ are calculated, where $D$ denotes the operator of the difference between the populations in the single-particle modes defined at the begining of Sec. V. For simplicity, the discussion is limited to the case of degenerate modes, i.e. $\varepsilon = 0$. The results are plotted versus time in units of $t_0 \equiv \frac{2\pi \hbar}{g_{1}}$ in Fig.~\ref{fig:Fig.9} for $N=200$ and different values of $g_{1}$ indicated in the figure for the rows (a-f). For values $0 \leq g_{1}/g_{0} < 1/3$ all atoms practically remain in the mode $\alpha_1$, i.e. self-trapping prevails. This is shown in row (a) for $g_{1}/g_{0} = 0.95 \times 1/3$, a value quite close to the critical value $g_{1}/g_{0} = 1/3$. From left to right $d_{\psi(t)}$, $\mathcal{F}[d_{\psi(t)}]$, $\Delta_{\psi(t)}d$, and $\mathcal{F}[\Delta_{\psi(t)}d]$ are shown with $\mathcal{F}[x(t)]$ denoting the Fourier spectrum of $x(t)$. The critical case $g_{1}/g_{0} = 1/3$ is shown in row (b). Self-trapping is now replaced by a rapid decay of $d_{\psi(t)}$ to zero with sharp revivals appearing at multiples of $t_0/4$, where all atoms alternately pile up in one of the modes $\alpha_i$. At these incidences naturally $\Delta_{\psi(t)}d$ reduces to zero, while for all other times a value on the order of $d_{\psi(t)}$ itself is attained, thus indicating strongly super-Poissonian fluctuations. The sharp resonances in $d_{\psi(t)}$ and $\Delta_{\psi(t)}d$ are reflected in the associated Fourier spectra through evenly spaced combs of harmonic frequencies. At an only slightly larger value $g_{1}/g_{0} = 1.05 \times 1/3$ in (c) these frequencies decohere thus giving rise to a seemingly irregular but nevertheless deterministic time-evolution. The situation remains similar over a wide range of ratios $g_{1}/g_{0}$ (cf. (d) and (e)) until in (f) the maximally possible value $g_{1}/g_{0} = 1$ is reached, where the dynamics is analogue to the case (b) with the only difference of a threefold shorter time scale $t_0$. The transition from the self-trapping regime to the pair tunneling regime at the critical ratio $g_{1}/g_{0} = 1/3$ sharpens with increasing particle number $N$ such that for $N \rightarrow \infty$ a non-equilibrium  phase transition occurs \cite{Hey:18}, while for the ground state no notable change arises here, as seen in Figs.~\ref{fig:Fig.4} - \ref{fig:Fig.7}.

\begin{figure}
\includegraphics[scale=0.3, angle=0, origin=c]{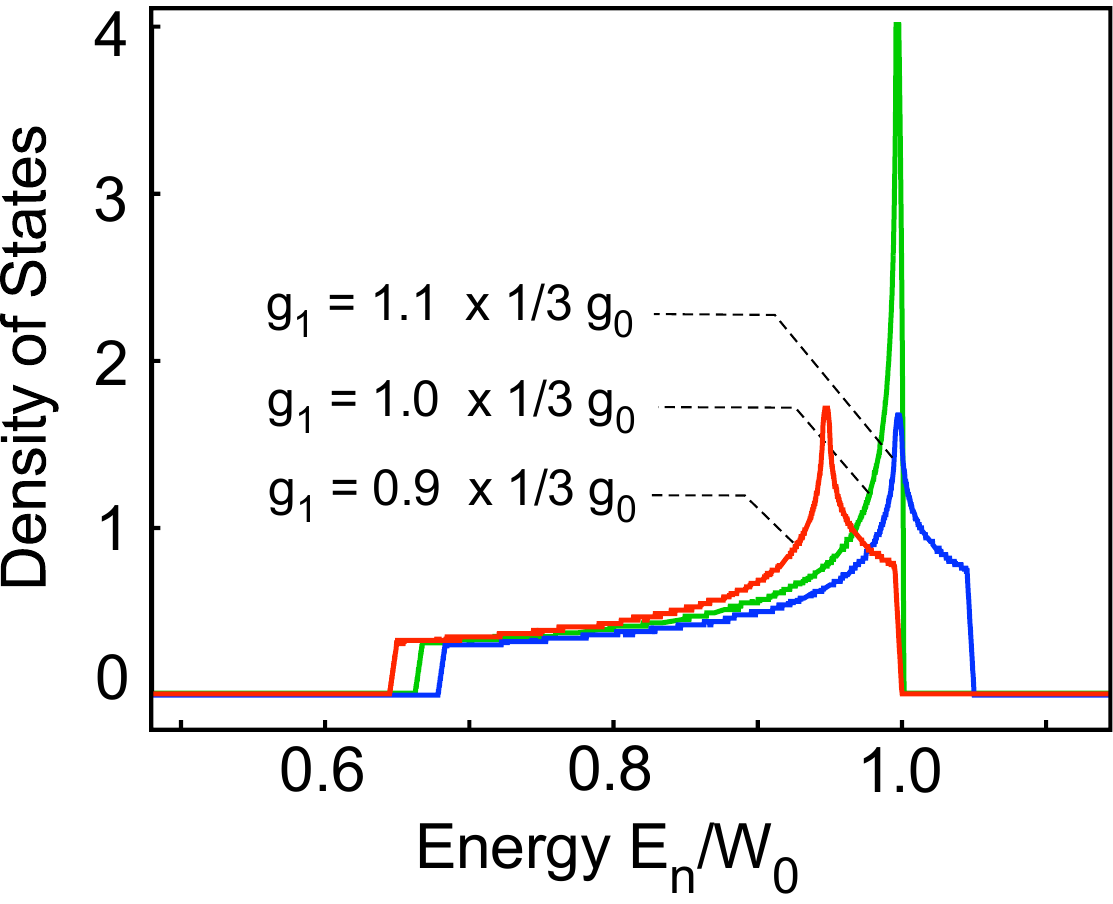}
\caption{The density of states is plotted for three indicated values of $g_{1}/g_{0}$ with $N=4000$ and $\varepsilon = 0$.}
\label{fig:Fig.10}
\end{figure}

To understand the peculiarity of the transition point one may revisit the structure of the eigenvalues in Fig.~\ref{fig:Fig.1}(a). One recognizes a resonance in the density of states that linearly increases from $\varepsilon_n = 1/2$ to $\varepsilon_n = 1$ as $g_{1}/g_{0}$ is tuned from zero to $\frac{1}{3}$. At this point a kink is observed, i.e., for $g_{1}/g_{0}$ in the interval $[\frac{1}{3},1]$ the energy of this resonance remains constant. This is more quantitatively seen in Fig.~\ref{fig:Fig.10}, where the density of states for $N=4000$ particles is plotted for $g_{1} = 0.9 \times \frac{1}{3} g_{0}$, $g_{1} =  \frac{1}{3} g_{0}$, and $g_{1} = 1.1 \times \frac{1}{3} g_{0}$, i.e. on both sides and exactly at the transition point. At the transition point the resonance of the density of states falls together with the upper edge of the energy spectrum. For $g_{1} <  \frac{1}{3} g_{0}$ the initial state $\ket{\psi(0)}$, with all atoms prepared in the same mode, energetically lies at the upper edge of the energy spectrum well above the resonance of the density of states, such that only a few eigenvectors are available that $\ket{\psi(0)}$ can be composed of. More specifically, $\ket{\psi(0)}$ is itself quite close to an eigenvector and therefore cannot significantly evolve in time. Hence, the phenomenon of self-trapping. If $g_{1} >  \frac{1}{3} g_{0}$, the initial state $\ket{\psi(0)}$ falls on the resonance of the density of states, i.e. many eigenstates are available to contribute to its composition, which enables the dramatic change of its dynamical properties.

\section{Experimental Implementation}
\begin{figure}
\includegraphics[scale=0.45, angle=0, origin=c]{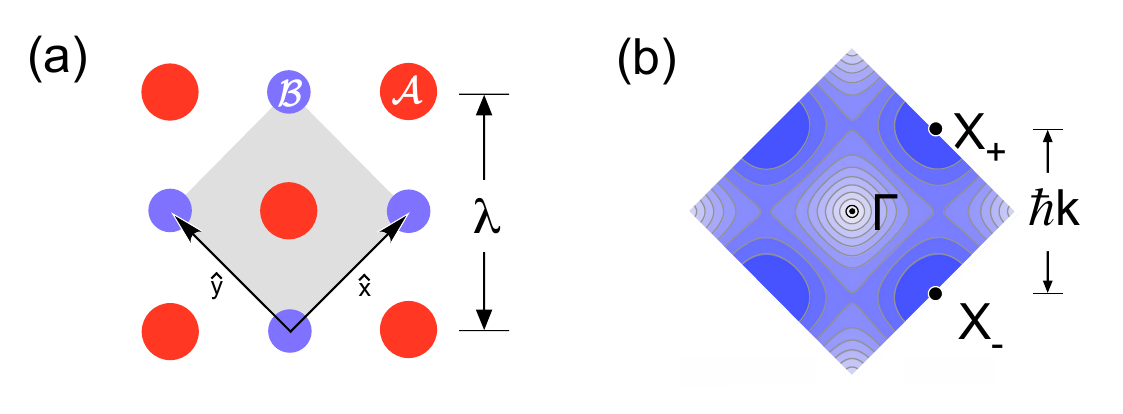}
\caption{(a) Sketch of the lattice geometry with deep A-sites and shallow B-sites. The unit cell is shown by the gray rectangle. (b) The second Bloch-band of the lattice in (a) is plotted across the first Brillouin zone with the two inequivalent energy minima at $X_{\pm}$ highlighted. Blue denotes low and white denotes high energy. The optical wavelength for generating the lattice is denoted $\lambda$ and $k=2\pi/\lambda$.}
\label{fig:Fig.11}
\end{figure}

This section begins with a brief introduction of an experimental platform that approximately implements the model Hamiltonian of Eq.~\ref{eq:Hamiltonian}. More detailed descriptions are found in Refs.~\cite{Oel:13, Koc:16}. The centrepiece of the experimental realization is a two-dimensional bipartite square optical lattice with the third dimension confined by a harmonic potential (with 40 Hz vibrational frequency), providing shallow and deep potential wells arranged as the black and white fields of a chequerboard, as sketched in Fig.~\ref{fig:Fig.11}(a). The second band of this lattice possesses two inequivalent local minima at two high symmetry points (denoted $X_{+}$ and $X_{-}$ located at the edge between the first and second Brillouin zones, as illustrated in Fig.~\ref{fig:Fig.11}(b). The experimental set-up allows one to tune the potential energy difference $\varepsilon$ of the $X_{\pm}$-points in quasi-momentum space and the relative energy difference $\Delta V$ between the deep and shallow wells in configuration space. As detailed in Refs.\cite{Oel:13, Koc:16} a long-lived BEC of rubidium atoms can be formed in the second band sharing both potential condensation points $X_{\pm}$. The Bloch functions  $\psi_{\pm}$ associated with $X_{\pm}$ take the role of the two single-particle mode functions at the basis of the model Hamiltonian of Eq.~\ref{eq:Hamiltonian}. A numerical band calculation allows one to determine the band structure, $\psi_{\pm}$, and the integrals $\rho_i, i \in \{0,1,2,3\}$, defined below Eq.~\ref{eq:Hamiltonian}, for arbitrary values of $\varepsilon$ and $\Delta V$. The Bloch functions are composed of local $s$-orbitals in the shallow wells and local $p_x$- and $p_y$-orbitals in the deep wells. Tuning of $\Delta V$ allows one to tune the fractions of atoms residing in the shallow and deep wells $\nu_{s}$ and $\nu_{p}$, respectively, where $\nu_{s}$ and $\nu_{p}$ are normalized to satisfy $\nu_{s}+\nu_{p}=1$. The collision parameters $g_i = g \rho_i$ can be numerically determined as functions of $\nu_{p}$.
 
Approximating the Bloch functions in terms of local $s$- and $p$-orbitals, simple analytical expressions of the collision parameters $g_i$ as functions of $\nu_{p}$ can be obtained. With the primitive vectors $\hat x$ and $\hat y$ from Fig.~\ref{fig:Fig.11}(a) and the lattice constant $a\equiv \frac{\lambda}{\sqrt{2}}$ one may write
\begin{eqnarray}
\label{eq:Bloch1}
\psi^{\pm}(x,y,z) = \qquad\qquad\qquad\qquad\qquad\qquad\qquad\qquad \\
\frac{s_z(z)}{\sqrt{M}} \sum_{n,m} \bigg(  (-1)^{[(n+m)\pm(n-m)]/2} \sqrt{\nu_s} \, s_{n,m}(x,y) \nonumber \\
+ (-1)^{[(n+m)\pm(m-n)]/2} \sqrt{\nu_p} \, p^{\pm}_{n,m}(x,y) \bigg)  \, , \nonumber  
\end{eqnarray}
where the sum extends over $M$ unit cells with
\begin{eqnarray}
\label{eq:Bloch2}
s_{n,m}(x,y) &\equiv& s(x-n a,y-m a) \nonumber  \\
p^{\pm}_{n,m}(x,y) &\equiv& p^{\pm}(x-[n-\frac{1}{2}]\, a,y-[m-\frac{1}{2}] \,a) \nonumber 
\end{eqnarray}
and $s(x,y)$ and $p^{\pm}(x,y)$ denoting the real-valued Wannier-functions associated with $s$- and $p^{\pm}$-orbitals, respectively, where $p^{\pm} \equiv p_{\frac{1}{2}[(x+y)\pm(x-y)]}$. 
Here, $s_z(z)$ denotes the ground state wave function of the harmonic oscillator trap potential with respect to the $z$-direction with the radius $\sigma_z \equiv \left(\int{dz |s_z(z)|^4}\right)^{-1}$ and the normalization relations $1 = \int{dxdy |s(x,y)|^2} = \int{dxdy |p^{\pm}(x,y)|^2} = \int{dz |s_z(z)|^2}$. With the approximation that orbitals in different lattice sites have negligible overlap one obtains the collision overlap integrals 
\begin{eqnarray}
\label{eq:Bloch3}
\rho_0 &=& \frac{1}{M \sigma_z}\bigg(\nu_{s}^2 \int dxdy |s(x,y)|^4 \nonumber \\
&+& \nu_{p}^2 \int dxdy |p^{\pm}(x,y)|^4 \bigg) \\
\rho_1 &=& \rho_2 = \frac{1}{M \sigma_z} \bigg(\nu_{s}^2 \int dxdy |s(x,y)|^4 \nonumber \\
&+&  \nu_{p}^2 \int dxdy |p^{+}(x,y)|^2 |p^{-}(x,y)|^2  \bigg) \nonumber \\
\rho_3 &=& \rho_4 = 0 \nonumber
\end{eqnarray}
Finally, applying a harmonic approximation for the lattice wells, $s(x,y) = s_{1D}(x)s_{1D}(y)$, $p_x(x,y) = $ $p_{1D}(x) s_{1D}(y)$, and $p_y(x,y) = s_{1D}(x) p_{1D}(y)$ with $s_{1D}(x) =$ $\sigma^{-1/2} \pi^{-1/4} e^{-x^2/2\sigma^2}$ and $p_{1D}(x) = \sigma^{-3/2} \pi^{-1/4} \sqrt{2}\, x \, e^{-x^2/2\sigma^2}$, one arrives at the simple expressions 
\begin{eqnarray}
\label{eq:Bloch4}
g_0 = g_{00} [(1-\nu_{p})^2 + \frac{3}{4}\nu_{p}^2] 
\nonumber \\ \\ \nonumber  
g_1 = g_{00} [(1-\nu_{p})^2 + \frac{1}{4}\nu_{p}^2]
\end{eqnarray}
with $g_{00} \equiv g/(2\pi M \sigma^2 \sigma_{z})$, which yields $g_0 - g_1 = g_{00} \, \nu_{p}^2 /2$. Note that $\nu_p$ is defined within the interval $[0,1]$, such that $1 - g_1/g_0$ lies in the interval $[0,2/3]$.

\begin{figure}
\includegraphics[scale=0.38, angle=0, origin=c]{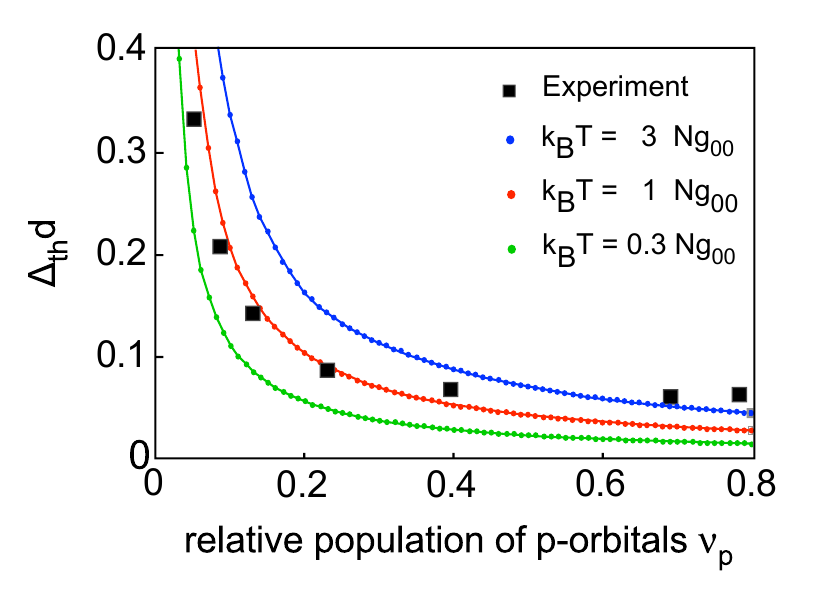}
\caption{The coloured disks connected by solid lines show the fluctuations $\Delta_{\textrm{th}}d$ versus $\nu_p$ calculated for $\varepsilon = 0$, $N=10^4$ particles, and $g_{00} = 10^{-5} E_{\textrm{rec}}$. The temperatures are in ascending order $k_B T = N g_{00} \times \{0.3,1,3\}$. The black squares show experimental data recorded with about $N \approx 4 \times 10^4$ at a temperature roughly estimated to be a few ten nanokelvin.}
\label{fig:Fig.12}
\end{figure}

With these preparations one can apply the general results for the model in Eq.~\ref{eq:Hamiltonian} to the present example. Previous work in Refs.\cite{Oel:13, Koc:16} has made use of the mean-field results in Fig.~\ref{fig:Fig.4}. Here, with the help of the full quantum model one may complement these considerations including fluctuations. In Fig.~\ref{fig:Fig.12}, experimental data for the fluctuations $\Delta_{\textrm{th}}d$ (black squares) are plotted versus $\nu_p$ and compared to calculations (red disks) using the quantum model described above. The calculations are performed for $\varepsilon = 0$, $N=10^4$ particles, $k_B T = N g_{00} \times \{0.3,1,3\}$ and $g_{00} = 10^{-5} E_{\textrm{rec}}$. This corresponds to the temperatures $\{3,10,30\}\,$nK. The data are obtained by conducting the following experimental protocol (cf. Ref.~\cite{Oel:13, Koc:16}): first, $\varepsilon=0$ is realized via precisely adjusting the intensities of all lattice beams. A BEC is loaded into the ground state of the lowest Bloch band. By rapidly ramping the chemical potential difference $\Delta V$ between A-sites and B-sites of the lattice (cf. Fig.~\ref{fig:Fig.11}(a)), the atoms are transferred into the second band. The chosen final value of $\Delta V$ determines the value of $\nu_p$. The atoms are then given several ten milliseconds time to condense with a significant condensate fraction populating the $X_{\pm}$-points. A momentum spectrum is obtained by a time-of-flight method and the number of atoms in each of the two lowest order Bragg resonances, corresponding to each condensation point, is recorded. To obtain reasonable statistics in the determination of $\Delta_{\textrm{th}}d$, for each data point several hundred momentum spectra are recorded and evaluated. The temperature can only be roughly estimated to be on the order of a few ten nK from that of the initial condensate in the lowest band. The particle number in the experiments is approximately $N = 4 \times 10^4$. According to Fig.~\ref{fig:Fig.12}, the observed fluctuations show the best agreement with the calculations for a temperature close to 10 nK. Increasing values of $\nu_p$ are associated with growing populations in the local $p$-orbitals of the deep wells, which increases band relaxation losses via binary collisions, where both atoms decay to the lower lying $s$-orbital. The associated heating of the remaining atoms should be responsible for the observed slight increase of the observed fluctuations for large $\nu_p$, which is not captured by the calculations.

\section{Conclusion}
In conclusion, a quantum model of bosons condensed in two orthogonal single-particle modes with flavour-changing contact interaction leads to a rich collective phenomenology. In a wide parameter range, the interaction induces coherence between the two single-particle modes in contrast to flavour conserving variants of contact interaction, which typically inhibit coherence. In a mean field description, two possible ground states arise associated with spontaneously broken time-reversal symmetry and a non-zero magnitude of angular momentum. In a full quantum treatment, a coherent (for zero or very small temperatures) or incoherent (for larger temperatures) superposition of the two possible mean field ground states arises. The fluctuations in the relative populations of both single-particle modes along certain paths in the phase diagram show universal scaling. An analysis in terms of the Penrose-Onsager criterion shows that the low temperature quantum ground state can be fragmented on the single-particle level, while pair correlations build up that maintain condensate character on the level of pairs. The non-equilibrium dynamics shows a sharp transition between a self-trapping and a pair-tunneling regime. The model captures central aspects of the physics of atoms condensed in the two inequivalent band minima of the second Bloch band of a bipartite optical square lattice. The physics of double wells with interaction induced pair tunneling and the collision physics in near degenerate $p_x$- and $p_y$-orbitals turns out to be intimately related.

\section*{Acknowledgements} 
This work was partially supported by DFG-SFB925 and the Hamburg Centre of Ultrafast Imaging (CUI). Max Hachmann is acknowledged for his expert advice on accelerating parts of the numerical code, and Claus Zimmermann, W. Vincent Liu, and Cristiane Morais Smith for useful discussions.


\begin{thebibliography}{21}

\bibitem{Leg:01}
A. J. Leggett, Rev. Mod. Phys. {\bf 73}, 307-356 (2001).

\bibitem{Pet:02}
C. J. Pethick and H. Smith, Bose-Einstein Condensation in Dilute Gases, Cambridge University Press (2002).

\bibitem{Pit:03}
L. P. Pitaevskii and S. Stringari, Bose-Einstein Condensation, Oxford University Press (2003).

\bibitem{Sme:97}
A. Smerzi, S. Fantoni, S. Giovanazzi, and S. R. Shenoy, Phys. Rev. Lett. {\bf 79}, 4950 (1997).

\bibitem{Mil:97}
G. J. Milburn, J. Corney, E. M. Wright, and D. F. Walls, Phys. Rev. A {\bf 55}, 4318 (1997).

\bibitem{Gri:98}
M. Grifoni and P. H{\"a}nggi, Phys. Rep. 304, 229 (1998).

\bibitem{Dal:12}
B. J. Dalton and S. Ghanbari, J. Mod. Optics, {\bf 59}, 287-353 (2012).

\bibitem{Noz:95}
P. Noz$\grave{\textrm{e}}$ires, \textit{in Bose-Einstein Condensation}, edited by A. Griffin, D. W. Snoke, and S. Stringari, Cambridge University press (1995).

\bibitem{Ruo:98}
J. Ruostekoski and D. F. Walls, Phys. Rev. A {\bf 58} , R50 (1998).

\bibitem{Kal:01}
G. Kalosakas and A. R. Bishop, Phys. Rev.  A {\bf 65}, 043616 (2002).

\bibitem{Mue:06}
E. J. Mueller, T.-L. Ho, M. Ueda, and G. Baym, Phys. Rev. A {\bf 74}, 033612 (2006).

\bibitem{Sal:07}
A.N. Salgueiro, A.F.R. de Toledo Piza, G. B. Lemos, R. Drumond, M.C. Nemes, and M. Weidem{\"u}ller, Eur. Phys. J. D {\bf 44}, 537-540 (2007).

\bibitem{Rap:12}
K. Rapedius, J. Phys. B: At. Mol. Opt. Phys. {\bf 45} 085303 (2012).

\bibitem{Bac:98}
S. Backhaus, R. W. Simmonds, A. Loshak, J. C. Davis, and R. E. Packard, Nature {\bf 392}, 687-690 (1998).

\bibitem{Alb:05}
M. Albiez, R. Gati, J. F{\"o}lling, S. Hunsmann, M. Cristiani, and M. K. Oberthaler, Phys. Rev. Lett. {\bf 95}, 010402 (2005).

\bibitem{Lev:07}
S. Levy,  E. Lahoud, I. Shomroni, and J. Steinhauer, Nature {\bf 449}, 579-583 (2007).

\bibitem{Foe:07}
S.  F{\"o}lling,  S.  Trotzky,  P.  Cheinet,  M.  Feld,  R.  Saers,  A.  Widera,  T.  M{\"u}ller  and  I. Bloch, Nature {\bf 448}, 1029 (2007).

\bibitem{Abb:13}
M. Abbarchi, A. Amo, V. G. Sala, D. D. Solnyshkov, H. Flayac, L. Ferrier, I. Sagnes, E. Galopin, A. Lema\^{i}tre, G. Malpuech, and J. Bloch, Nat. Phys. {\bf 9}, 275 (2013).

\bibitem{Lia:09}
J.-Q. Liang, J.-L. Liu, W.-D. Li, and Z.-J. Li, Phys. Rev. A {\bf 79}, 033617 (2009).

\bibitem{Bad:09}
P. Bader and U. R. Fischer, Phys. Rev. Lett. {\bf 103} 060402 (2009).

\bibitem{Cao:12}
H. Cao and L.B. Fu,  Eur. Phys. J. D (2012) 66: 97.

\bibitem{Jas:12}
P. Jason and M. Johansson, Phys. Rev. A {\bf 85}, 011603(R) (2012).

\bibitem{Fis:13}
U. R. Fischer and B. Xiong, Phys. Rev. A {\bf 88}, 053602 (2013).

\bibitem{Zhu:15}
Q. Zhu, Q. Zhang, and B. Wu, J. Phys. B: At. Mol. Opt. Phys. {\bf 48}, 045301 (2015).

\bibitem{Rub:17}
D. Rubeni, J. Links, P. S. Isaac, and A. Foerster, Phys. Rev. A  {\bf 95}, 043607 (2017).

\bibitem{Agb:18}
D. Agboola, P. S. Isaac, and J. Links, J. Phys. B: At. Mol. Opt. Phys. {\bf 51} 145301 (2018).

\bibitem{Pen:56}
O. Penrose and L. Onsager, Phys. Rev. {\bf 104} 576 (1956).

\bibitem{Isa:05}
A. Isacsson and S. M. Girvin, Phys. Rev. A {\bf 72}, 053604 (2005).

\bibitem{Liu:06}
W. V. Liu and C. Wu, Phys. Rev. A {\bf 74}, 013607 (2006). 
 
\bibitem{Cai:11}
Z. Cai and C. Wu, Phys. Rev. A {\bf 84}, 033635 (2011).

\bibitem{Oel:13}
M. \"{O}lschl\"{a}ger, T. Kock, G. Wirth, A. Ewerbeck, C. Morais Smith, and A. Hemmerich, New J. Phys. {\bf 15}, 083041 (2013).

\bibitem{Li:16}
X. Li and W. V. Liu, Rep. Prog. Phys. {\bf 79}, 116401 (2016).

\bibitem{Koc:16}
T. Kock, C. Hippler, A. Ewerbeck, and A. Hemmerich, J. Phys. B: At. Mol. Opt. Phys. {\bf 49}, 042001 (2016).

\bibitem{Hey:18}
Markus Heyl, Rep. Prog. Phys. {\bf 81}, 054001 (2018).

\end{thebibliography}
\end{document}